\documentclass[sigconf]{acmart}

\usepackage{lipsum}  

\usepackage{wrapfig}

\usepackage{booktabs}                  
\usepackage{lipsum}                    
\usepackage{mwe}                       

\usepackage{cellspace}

\usepackage{enumerate}
\usepackage{enumitem}

\usepackage{makecell} 
\usepackage{booktabs} 

\usepackage{amssymb} 

\usepackage{booktabs} 
\usepackage{tabularx} 
\usepackage[utf8]{inputenc}
\usepackage{geometry}
\usepackage{multicol}

\usepackage{xspace,xpunctuate}

\newcommand{\etal}{{~\textit{et al.}}}

\usepackage{tikz}  
\usetikzlibrary{shadings}
\definecolor{cback}{HTML}{E9ECEF}
\definecolor{cframe}{HTML}{495057}

\newcommand*\circled[1]{\tikz[baseline=(char.base)]{
    \node[shape=rectangle,rounded corners=1.5pt,fill=cback,text=black,draw=cframe,inner sep=1pt] (char) {#1};}}

\newcommand{\ac}[1]{#1}
\newcommand{\rc}[1]{#1}

\newcommand{\fc}[1]{#1}

\usepackage[nameinlink,capitalise]{cleveref}
\usepackage{array}

\AtBeginDocument{%
  \providecommand\BibTeX{{%
    Bib\TeX}}}

\copyrightyear{2025}
\acmYear{2025}
\setcopyright{cc}
\setcctype{by}
\acmConference[CHI '25]{CHI Conference on Human Factors in Computing Systems}{April 26-May 1, 2025}{Yokohama, Japan}
\acmBooktitle{CHI Conference on Human Factors in Computing Systems (CHI '25), April 26-May 1, 2025, Yokohama, Japan}
\acmDOI{10.1145/3706598.3714265}
\acmISBN{979-8-4007-1394-1/25/04}

\begin{document}

\title{\textit{TangibleNet}: Synchronous Network Data Storytelling through Tangible Interactions in Augmented Reality}

\author{Kentaro Takahira}
\orcid{1234-5678-9012}
\affiliation{%
  \institution{The Hong Kong University of Science and Technology}
  \city{Hong Kong}
  \country{China}
}
\email{ktakahira@connect.ust.hk}

\author{Wong Kam-Kwai}
\affiliation{
  \institution{The Hong Kong University of Science and Technology}
  \city{Hong Kong}
  \country{China}
}
\email{kkwongar@connect.ust.hk}

\author{Leni Yang}
\affiliation{
  \institution{The Hong Kong University of Science and Technology}
  \city{Hong Kong}
  \country{China}
}
\email{lyangbb@connect.ust.hk}

\author{Xian Xu}
\affiliation{%
  \institution{The Hong Kong University of Science and Technology}
  \city{Guangzhou}
  \country{China}
}
\email{xxubq@connect.ust.hk}

\author{Takanori Fujiwara}
\affiliation{
  \institution{Linköping University}
  \city{Linköping}
  \country{Sweden}
}
\email{takanori.fujiwara@liu.se}

\author{Huamin Qu}
\affiliation{%
  \institution{The Hong Kong University of Science and Technology}
  \city{Hong Kong}
  \country{China}
}
\email{huamin@cse.ust.hk}

\renewcommand{\shortauthors}{Takahira et al.}


\begin{abstract}
Synchronous data-driven storytelling with network visualizations presents significant challenges due to the complexity of real-time manipulation of network components. 
While existing research addresses asynchronous scenarios, there is a lack of effective tools for live presentations. 
To address this gap, we developed TangibleNet, a projector-based AR prototype that allows presenters to interact with node-link diagrams using double-sided magnets during live presentations. 
The design process was informed by interviews with professionals experienced in synchronous data storytelling and workshops with 14 HCI/VIS researchers. Insights from the interviews helped identify key design considerations for integrating physical objects as interactive tools in presentation contexts. The workshops contributed to the development of a design space mapping user actions to interaction commands for node-link diagrams. Evaluation with 12 participants confirmed that TangibleNet supports intuitive interactions and enhances presenter autonomy, demonstrating its effectiveness for synchronous network-based data storytelling.

\end{abstract}



\begin{teaserfigure}
  \centering
  \includegraphics[width=1\textwidth]{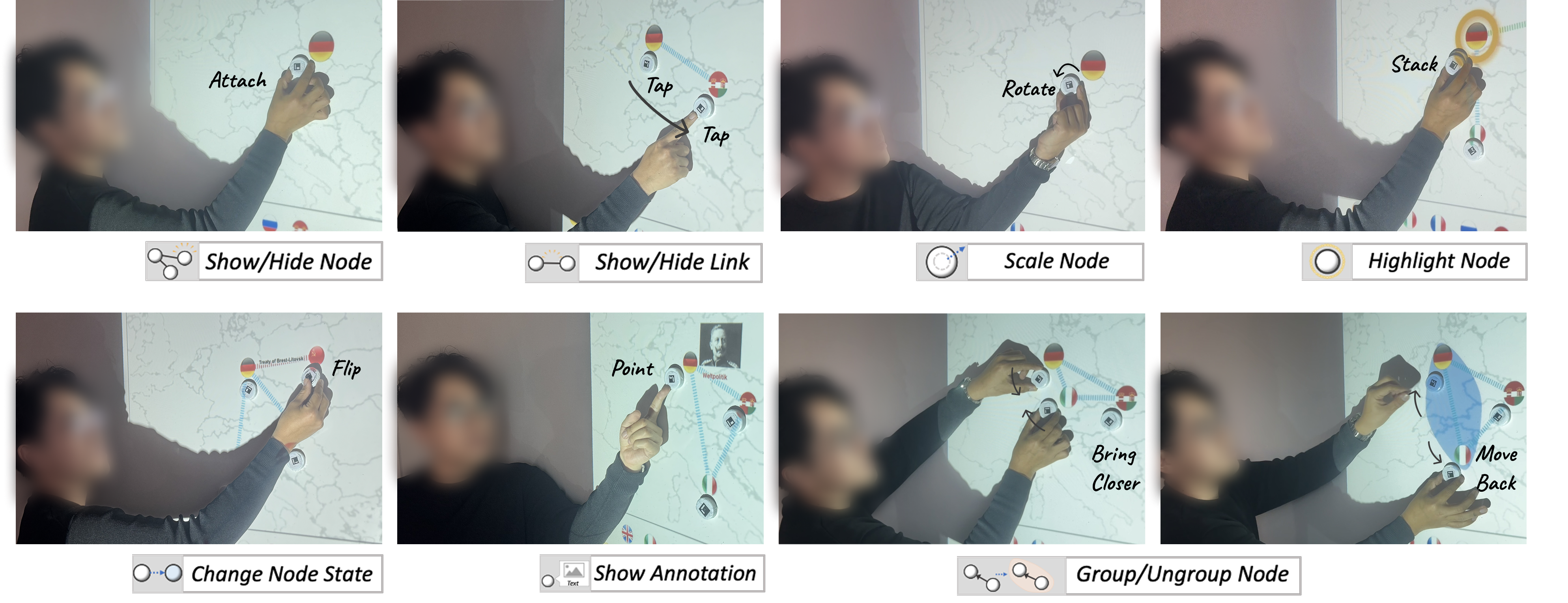}
  \caption{TangibleNet is a projector-based AR prototype for live data storytelling using network visualizations. Presenters interact with node-link diagrams through double-sided magnets and hand gestures. By leveraging the affordance of physical objects, TangibleNet enables quick, engaging interactions and provides an improvisational presentation experience.}
  \label{fig:teaser1}
  \Description{This teaser image illustrates the main visualization commands available in TangibleNet, a system for interacting with network visualizations during live presentations using physical objects. It highlights the commands and the corresponding user actions required to perform them. Further details about these actions and commands can be found in Figure 6.}
\end{teaserfigure}


\maketitle
\section{Introduction}
Networks are prevalent in data-driven storytelling, modeling phenomena in various fields such as international relations, ecosystems, and social networks~\cite{newman2018networks}. 
Significant efforts have been made to improve the communication of network data through various formats, including infographics~\cite{SpritzerStaticNet}, comics~\cite{kimDataToon, bashDynamicGraphComic, DGComics}, and animations~\cite{graphDiaries, DiffAni}.
However, research in network-based data storytelling predominantly focuses on asynchronous scenarios, where audiences engage with the content individually, without interaction or live participation from narrators~\cite{kosara2013storytelling}.
This emphasis \fc{does not extend to} synchronous scenarios, where narrators guide the audience through the narratives in real-time.

Synchronous data storytelling has become increasingly common and \rc{important} in various contexts. 
\rc{It is frequently adopted in organizational decision-making meetings~\cite{dimara2021unmet, FromJamtoRecital} (e.g., executives illustrate business strategies with corporate networks) and public communication~\cite{kosara2013storytelling} (e.g., news anchors explain international affairs to audiences).}
Successful synchronous presentations require careful \rc{planning and coordination, such as timing visual aids to the narrative pace, complementing the stories with different} communication modalities (e.g., voice and gestures), and fostering dynamic interactions between the presenter and the audiences \cite{Chironomia, Elastica, BodyDrivenTalk, RealityTalk}.
A prominent example is Hans Rosling's presentations, where he synchronizes gestures and body postures, such as pointing out patterns and tracing trends, with animated visualizations to guide the audiences' attention~\cite{rosling2007best, rosling2011countries}.
Additionally, Rosling incorporated physical objects to interact with charts and engage the audience. 
For instance, he once used a meter-long teaching stick in a presentation~\cite{rosling2015ignorant}, eliciting laughter from the audience.
In his talks about global population growth~\cite{rosling2016global, rosling2016world}, he used boxes to represent populations and demonstrated the population dynamics by stacking and unstacking the boxes\ac{, making abstract data more tangible and understandable.}

\rc{Creating dynamic narratives similar to Rosling's presentations in traditional slideshow software (e.g., PowerPoint) requires preparing multiple slides for each state of the charts, along with carefully coordinated transitions and animations.}
This \rc{process is time-consuming and lacks support for the gesture and physical object-based interactions, which can provide advantages like enhanced understanding~\cite{dargue2019our} and engagement~\cite{Chironomia}.}
\rc{Prior} research has explored innovative authoring tools that enable presenters to map gestures and postures to control \rc{visual} elements in augmented presentations~\cite{BodyDrivenTalk, Chironomia, visConductor, Elastica}.
\ac{However, these tools do not support network-based data storytelling.}
In network-based data storytelling, presenters often use node-link diagrams~\cite{networkNerratives} and need to manipulate various visual components (e.g., nodes, \rc{links}, and annotations) and their attributes (e.g., color, position, and size) at both individual and group levels~\cite{bashDynamicGraphComic}. 
\ac{Relying solely on gestures to map all possible actions can lead to overly complex gestures, making them difficult to remember, prone to errors, and vulnerable to recognition failures.
The limitations have been found in evaluations of systems that rely solely on gesture-based interactions with visualizations~\cite{Chironomia}.}

This work addresses the above two-fold challenge---the lack of augmented presentation support for networks and the complexity of gesture-based interactions. To overcome these, we propose an approach where presenters use physical objects to interact with network visualizations during their presentations.
Physical objects offer diverse affordances, such as direct manipulation and spatial interaction, that simplify complex tasks through \rc{familiar} interactions~\cite{grappableUI, TUItextbook}. For instance, prior studies have used physical objects like cubes, spheres, and sticks, employing actions such as flipping, stacking, and combining to streamline complex visual manipulations~\cite{bricks, DataCube, sphere, stick}.
Leveraging these affordances can resolve the complexity of gesture-based interactions and enhance the intuitiveness of interacting with network visualizations.
To illustrate this concept, we developed TangibleNet, a portable, projector-based augmented reality (AR) prototype.
TangibleNet enables \ac{presenters} to interact with network visualizations projected onto a whiteboard using double-sided magnets. 

To inform the system design, we first interviewed five professionals, specifically news anchors, whose roles involve narrating data stories synchronously with visuals.
Despite their extensive expertise in data communication, they have been largely overlooked in visualization research. 
While our prototype targets a broader audience beyond news anchors, we aimed to uncover insights \ac{on effective real-time communication} not yet explored in the literature.
We then conducted workshops with 14 HCI/VIS researchers to explore interaction methods for network visualizations using physical objects. 
These workshops led to a design space characterized by three key dimensions: 1) Interaction Command, 2) Primary Modality, and 3) Multiplexity of Physical Objects. 
Insights from both studies reinforced the untapped potential of using physical objects and informed design considerations for synchronous network data storytelling.
Building on these findings, we developed TangibleNet and evaluated it with 12 participants. 
\rc{Most participants provided positive feedback on 
the naturalness of interactions, the engaging delivery process, and the enhanced sense of autonomy during presentations. 
We synthesized insights from the prototype and user feedback to propose design implications for future systems supporting physical interactions in synchronous data storytelling. }
In summary, our contributions are three-fold:
\begin{itemize}[noitemsep,topsep=0pt,label=$\diamond$, leftmargin=*]
\item \textbf{Novel Scenario}:
We introduce synchronous network-based data storytelling, identifying key communication elements and system requirements through interviews with previously overlooked data communicators--news anchors (N=5).

\item \textbf{Design Space}:
We propose a framework for interacting with node-link diagrams using physical objects based on insights from a workshop with VIS/HCI researchers (N=14).

\item \textbf{TangibleNet Prototype}:
We develop and evaluate TangibleNet, demonstrating how physical objects enable network visualization interactions in synchronous storytelling. We confirmed its effectiveness through user studies (N=12).
\end{itemize}
\section{Related Works}


Our research explores the intersection of synchronous presentation, network data storytelling, and physical objects for data visualization interactions.

\subsection{Synchronous Data-driven Storytelling}
Traditionally, data-driven storytelling has focused on producing asynchronous content \cite{dataStorySurveyHaotian}. However, as data-driven decision-making becomes increasingly prevalent from casual discussions to formal presentations, \rc{the need for synchronous} data-driven storytelling has grown significantly \cite{Chironomia, dataTV, FromJamtoRecital,visConductor}. This approach integrates multiple forms of communication, including speech, gestures, eye gaze, and physical or virtual props, to create more engaging and adaptable presentations \cite{GestureAsCommunication, CoordinationForPresentation,lund2007importance,engle2000toward}. As Kang\etal~\cite{CoordinationForPresentation} pointed out, effective presentations require the seamless coordination of gestures, language, and props. Many systems have been developed to support this integration.

\rc{Saquib\etal~\cite{BodyDrivenTalk}} investigated body-driven graphics, where pre-designed visuals are mapped to specific body parts and adjusted in response to the presenter's movements. 
RealityTalk \cite{RealityTalk} utilizes a keyword-matching system to link spoken words with graphical elements. 
This system displays predefined graphics in real-time when specific keywords are recognized in speech, and these graphics can then be manipulated through hand gestures. Elastica \cite{Elastica} tackles the issue of recognition errors and presenter mistakes by enabling the dynamic adjustment of predefined graphic animations using both speech and gestures. This system allows presenters to define visual effects dynamically by combining body movements and spoken input. 

The effective communication of data-driven insights requires purpose-built systems, as general augmented presentation tools are inadequate for this task \cite{Chironomia}. In an interview study, Brehmer\etal~\cite{FromJamtoRecital} found that synchronous data storytelling can range from interactive, jam session-style presentations with flexible data visualization interactions to recital-style presentations with minimal audience engagement. The system requirements differ significantly across these formats, leading the authors to propose prototypes for various needs.
In their early exploration, Lee\etal~\cite{LeeSketchStory} introduced SketchStory, a system for dynamic chart creation, annotation, and filtering via touch and pen on a wall display. However, it requires \rc{pre-registering visual orders to minimize mode switching} and interaction complexity, limiting the flexibility essential for synchronous data storytelling~\cite{FromJamtoRecital, amini2018evaluating}. Additionally, while focused on visualization creation, its support for fine-grained manipulation is limited, making it challenging to synchronize speech with visuals closely.
Hall\etal~\cite{Chironomia} introduced Augmented Chironomia, a system designed \rc{for remote presentations} that enables gesture-based control of visualizations, supported by an authoring tool~\cite{visConductor}. 
This system overlays the presenter's webcam feed with interactive charts that can be manipulated in real-time through gestures. 
\ac{While this approach works well for these types of visualizations, node-link diagrams pose distinct challenges. Interacting with node-link diagrams requires managing a diverse set of visual components (e.g., nodes, links, and annotations) and their attributes (e.g., color, position, and size) at both individual and group levels \cite{taskTaxonomyGraphBongshin, kimDataToon}.}
Relying solely on body gestures for all interactions \rc{can lead to overly complex gesture mappings}, making them difficult to memorize, error-prone, and prone to misinterpretation. \ac{While combining speech and hand gestures has been proposed as a solution to expand the range of commands, using imperative voice commands can feel awkward and potentially distract the audience~\cite{srinivasancombining}.}

\subsection{Storytelling with Networks}
Networks play a pivotal role in data-driven storytelling because of their structural flexibility and the \rc{clear representation} they provide through node-link diagrams. They are widely used in various media, such as journalism, data videos, and data comics, to represent a wide range of topics, including interpersonal relationships, international relations, and ecosystems~\cite{newman2018networks, PatternsinAwardWinning, networkSurveyofSurvey, NarratingNetworks, kimDataToon, bashDynamicGraphComic}. 
Significant efforts have been made to develop tools that effectively communicate network data stories.

\rc{Spritzer\etal~\cite{SpritzerStaticNet} developed a system to enhance node-link diagrams by allowing users to modify visual attributes and layouts, facilitating the creation of more communicative visualizations.}
Similarly, \ac{Romat\etal~\cite{ExpressiveAuthoringNodeLink} proposed an interactive system that allows users to adjust visual attributes of multivariate network visualizations. 
Complementing these efforts, computational methods have also been introduced to support the effective communication of insights derived from network data analysis. Fujiwara\etal~\cite{FUJIWARA2018213} presented a system that automatically composes concise visual summaries of network analysis provenance, aiding in the sharing and recalling of analysis processes and results. Chen\etal~\cite{CalliopeNet} proposed Calliope-Net, a system designed to automatically extract and annotate salient topological features in node-link diagrams to produce visually appealing fact sheets of networks. 
While these tools enhance the aesthetics and clarity and effectively summarize key insights, they do not address the dynamic visual transitions important for storytelling~\cite{bashDynamicGraphComic}.} 

\ac{To effectively narrate changes in network data storytelling, data comics have emerged as a compelling format.}
Bach\etal~\cite{bashDynamicGraphComic} 
identified key design factors for representing dynamic networks in data comics, including various graph elements, component types, visual representations, and narrative patterns.
Building on this foundation, Kim\etal~\cite{kimDataToon} developed DataToon, an interactive authoring tool for creating network data stories in data comics.
This tool allows users to issue various commands through pen and multi-touch inputs via mode changes.
Additionally, Kim\etal~\cite{DGComics} developed a semi-automatic authoring tool designed explicitly for crafting data comics. 
\rc{These studies offer valuable insights into the key elements of network data storytelling, including essential network components, types of changes, visual encoding techniques, and narrative styles.} \fc{However, their focus is on developing tools and content for asynchronous consumption, which differs from the design of spontaneous, easily executed interactions needed for synchronous storytelling.} 
Consequently, live presentations often depend on static screenshots of network visualizations arranged in slide decks or rely on pre-determined sequences to navigate various components (e.g., XMind~\cite{xmind}). 
These approaches are not only labor-intensive and difficult to update but also limit the presenter's ability to engage in spontaneous, real-time interactions with the visualizations. 
\ac{This limitation restricts the presenter's capacity to dynamically adapt to audience needs, reducing the opportunity to deliver a personalized, engaging, and interactive storytelling experience.}

\begin{table*}[t!]
\centering
\caption{Profiles of the five interviewed news anchors.}
\resizebox{0.9\textwidth}{!}{%
\begin{tabular}{>{\centering\arraybackslash}m{1cm}m{1cm}m{2cm}m{4cm}m{7cm}}
\Xhline{4\arrayrulewidth}
\textbf{ID} & \textbf{Gender} & \textbf{Experience} & \textbf{Areas of Specialization}  & \textbf{Example Data}  \\ \Xhline{3\arrayrulewidth}

A1 & Male & 10 years &  Economic Affairs; Social Issues; Criminal Justice  & 
$\diamond$ Corporate Dynamics \newline
$\diamond$ Electoral Influence Analysis \newline
$\diamond$  Accident Statistics\\ \hline

A2 & Male & 8 years  & Sports; Social Trends; Crime Reporting &  
$\diamond$ Sports Coverage \newline
$\diamond$ Accident Statistics \newline
$\diamond$ Crime Relationship Mapping \\ \hline

A3 & Female & 10 years & Public Administration; Economic Affairs & 
$\diamond$ Governance Metrics (e.g Administrative Statistics)\newline
$\diamond$ Economic Indicators (e.g Business Bankruptcy Rates and Job Vacancy Ratios)\\ \hline

A4 & Male & 11 years & Criminal Justice; Political Analysis; Economic Reporting  & 
$\diamond$ Electoral Metrics (e.g Vote Counts, Influence Analysis) \newline
$\diamond$ Crime Network\newline
$\diamond$ Political Affiliation Analysis \\  \hline

A5 & Male & 10 years & Sports; Educational Reporting; Social Trends & 
$\diamond$ Sports Metrics (e.g Sports Results and Statistics)  \newline
$\diamond$ Educational Statistics \newline
$\diamond$ Relations of Prominent Figures \\ \Xhline{3\arrayrulewidth}

\end{tabular}%
}
\label{tab:experts_list}
\end{table*}

\subsection{Interacting with Visualizations Using Physical Objects}
Humans excel at sensing and manipulating physical objects, making them an effective medium for intuitive, low-effort interactions~\cite{TangibleBit, TUItextbook, IshiiTUI}.
Physical objects also offer spatial multiplexing, allowing users to control multiple virtual elements through physical arrangements~\cite{{grappableUI}}. 
Consequently, \rc{physical interactions} have attracted growing interest, particularly in augmented reality~\cite{affordanceBased, AnnexingReality}. 
Several studies have explored using simple geometric objects (e.g., cubes, cylinders, spheres) for virtual interaction~\cite{sphere, bricks, cylinder, stick}, leveraging actions like rotating, relocating, stacking, and tapping~\cite{DataCube, TangibleTouch, PaperVis}. These simple shapes are versatile and applicable to a wide range of interaction designs.
By aligning with users' real-world experiences, these interactions help reduce learning effort and enhance usability~\cite{TUItextbook}.

Recent research has explored the use of physical objects in data analytics, \rc{emphasizing their potential to perform interaction commands (e.g., selecting, filtering, highlighting) across various visualization types and contexts. }
Ens~\etal~\cite{uplift} introduced Uplift, which integrates tangible widgets with AR for energy analysis, using physical models and the bespoke slider to support tasks such as annotation and filtering in collaborative settings.
Satriadi\etal~\cite{tangibleGlobe} explored tangible globes for geospatial visualization, \rc{leveraging affordances like rotation and tapping.}
Satriadi\etal~\cite{activeProxyDash} also proposed the Active Proxy Dashboard, enabling interaction with physical scale models for selection and filtering.
Cordeil\etal~\cite{embodiedAxis} developed Embodied Axis, \rc{a system where tangible arms represent data axes. Users can spatially combine these axes and manipulate levers for selection and authoring.} 
Suzuki\etal~\cite{shapebot} explored shape-changing swarm interfaces, where users physically manipulate robots to construct visualizations.

Additionally, prior studies explored the use of data physicalization to connect physical and digital representations, allowing users to interact with digital data through tangible means~\cite{SandraSurveyDataPhy}.
Veldhuis~\etal~\cite{Coda} developed a tangible scatterplot where users can manipulate tokens as data points to explore correlations, particularly in educational contexts.
\fc{Taher~\etal~\cite{physicallyDynamicBar} examined how people interact with physical bar charts for common tasks like filtering and annotation. 
Le Goc~\etal~\cite{CompositeUsingWheeled} introduced self-propelled micro-robots for interactive data manipulation.
Additionally, \ac{Bae\etal~\cite{NetworkPhyAuthor} employed physicalized networks to facilitate basic interactions with node-link diagrams, such as highlighting or filtering.} 
Jansen and Dragicevic~\cite{BeyondDesktop} proposed a conceptual framework for beyond-desktop interaction, which integrates tangible and embodied interactions into visualization models.}

Researchers have also examined everyday objects for interaction. 
Tong\etal~\cite{PaperVis} proposed a design space for paper-based interactions, using affordances of papers \ac{such as folding, flipping, and tilting for various visualization tasks like filtering, zooming, and authoring.} 
He\etal~\cite{DataCube} explored cubes of different sizes for interacting with spatiotemporal data in mixed reality.
While these studies address a range of visualization tasks and provide valuable insights into the rich affordances of physical objects, few specifically focus on network visualizations and do not consider their use in live data storytelling.

\ac{In summary,} \rc{previous research has yet to explore the potential of physical interactions in synchronous data-driven storytelling,} \ac{especially for network visualizations.} While the importance of physical objects in storytelling has been recognized~\cite{riche2018data} and demonstrated in compelling cases~\cite{rosling2014global,rosling2016global}, their application in network data storytelling remains largely unexamined. 








\begin{figure*}[t]
    \centering
    \includegraphics[width=0.95\linewidth]{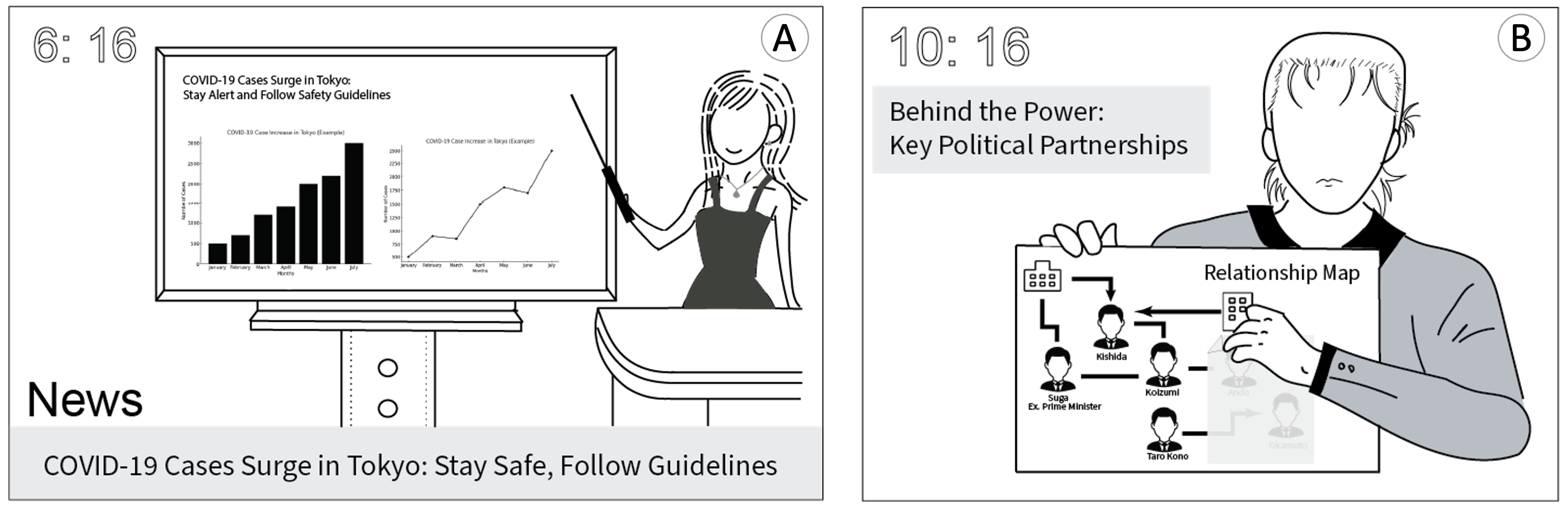}
    \caption{Illustrations of news anchors at work: (a) A3 explaining the rise in COVID-19 cases in Tokyo during a news segment. (b) A scene of a news anchor explaining the relationships between politicians using a physical board, as described by A4 and A5. The visuals are revealed in sync with the narration by removing stickers.}
    \label{fig:tvanchor}
    \Description{This figure contains illustrations drawn from interviews with a news anchor. The first illustration shows the anchor explaining trends in Tokyo’s COVID-19 cases. The second illustration shows the anchor using a physical board to present a political relationship map, where he is peeling stickers off the board to reveal information gradually.}
\end{figure*}

\section{Formative Study}
We first interviewed experienced news anchors to gain insights into their effective data communication practices and their views on using physical objects for synchronous data storytelling. 
These interviews helped identify the system requirements to support effective communication and the design considerations for integrating physical objects in presentation settings.
Despite their extensive expertise in conveying data to broad audiences in real-time, \ac{to the best of our knowledge,} news anchors have been largely overlooked by the visualization community. Although our approach of using physical objects to interact with network visualizations targets a broader audience, we aimed to uncover insights not yet covered \ac{in the existing literature~\cite{FromJamtoRecital, Chironomia, LeeSketchStory, Elastica}.}

\subsection{Interviews with News Anchors}
We conducted semi-structured interviews with five news anchors (A1-A5) who regularly engage in data storytelling using visuals. The participants (one female, four males) had 8 and 11 years of experience and were recruited via snowballing sampling, starting with a personal acquaintance of one author (\cref{tab:experts_list}). They work at TV stations in Japan, representing two different institutions \rc{and based at different broadcasting locations}. 
At the start of each interview, we outlined the project and obtained consent to record audio and take notes. We assured them that their identities would remain anonymous, their affiliated institutions would not be disclosed, and the data would be used exclusively for academic research purposes.
We asked about their current practices in explaining data with visual aids, challenges, key aspects of effective communication, and the nature of network data stories they presented. 
We also explored their interests, expectations, and concerns about using \rc{physical} objects in presentations. Each session lasted 90--120 minutes.

\subsection{Data Analysis}

The semi-structured interview results were saved as audio recordings and the interviewer's notes. These audio recordings were transcribed into text and then combined with the notes. We used an open coding procedure to analyze this data. 
One of the authors reviewed all records, assigning codes to data segments representing a single idea or concept.
The similar codes were then grouped, and each group was assigned a category. 
Three of the authors reviewed the groupings and categories, discussing their appropriateness until they agreed on the final coding results.

\subsection{Result}
The interview results confirmed the importance of interacting with visualizations in presentations, highlighted the potential of using physical objects for interaction, and provided insights into the challenges and design requirements of manipulating network visualizations with physical objects.

\subsubsection{Current Practices of Interaction with Visualization in Presentations}
All participants emphasized the importance of interaction with visuals during presentations for effective communication.
Common practices included touch interaction via multi-touch displays and manually notifying studio staff to proceed with slides, as noted by A1: \ac{\textit{``I just give them a nod to switch''}} \ac{(\cref{fig:tvanchor}-a)}. \rc{Additionally, participants} also used physical boards \rc{covered with removable stickers, which could be progressively taken off to reveal new information} (\cref{fig:tvanchor}-b).
All participants agreed on the effectiveness of gradually revealing the story through interaction, which aligns with previous research findings~\cite{FromJamtoRecital}.
This approach was \rc{particularly valued} for its flexibility, \rc{especially in integrating} real-time data (e.g., disaster updates) and accommodating ad-hoc \rc{changes in presentation content and timing}.
Additionally, A2-A4 \rc{further highlighted} \rc{how body movement enhances communication by adding a performative aspect to the presentation.} 
For example, A3 \ac{explained, \textit{``Using a pointer and body gestures to trace an upward trend in COVID case numbers really drives the point home''}} (\cref{fig:tvanchor}-a).
\rc{Similarly, A2 emphasized, \ac{\textit{``Removing stickers from a board to progressively reveal information engages the audience like uncovering mysteries''}}} (\cref{fig:tvanchor}-b).

\subsubsection{Presentation with Networks}
Participants explained the prevalence of presenting network data due to its structural versatility. They worked with various network data represented in node-link diagrams, such as relationships between political figures, international relations, inter-company networks, family trees, and crime connections. \ac{A4 noted, \textit{``Network diagrams let you map complex narratives, like showing alliances and conflicts between political figures and their parties during the last election.''}} These networks were typically composed of nodes, \rc{links}, groups, and captions (e.g., text or images) featuring up to 10 nodes\rc{, often represented with icons or images.} 
Participants adjusted visuals or revealed new elements as the story progressed. 
Unlike text or standard charts like bar or line graphs, node-link diagrams \rc{lack a set narrative direction (e.g., top-left to bottom-right)}, making it essential \rc{for presenters} to guide the audience's focus. 
Participants \rc{shared storytelling strategies, \textit{``starting with an overview, then zooming in on a specific node to explain a particular relationship, like showing intra-party dynamics after discussing inter-party relations,''} or \textit{``comparing multiple networks, such as contrasting political alliances,''} as noted by A4.}
\subsubsection{Opportunities for Using Physical Objects}
Participants expressed an interest in using physical objects during presentations \rc{and suggested that familiar, tangible objects would enable them} to focus on storytelling rather than system operations. The idea of using physical objects to represent network nodes was well-received.
A2 noted, \rc{\textit{``I've used magnets to represent players in sports broadcasts, and this idea feels natural and intuitive.''}}
\rc{Physical interactions viewed as a way to enhance audience engagement, transforming the presentation into a form of performance art.} 
\rc{As A3 put it, \textit{``Using props like TV shows do could make abstract data more understandable to the broader audience.''}}
A1 appreciated the simplicity of using familiar objects, stating, \rc{\textit{``Familiar objects are easier to handle than having to remember complex gestures or touch commands.''}}

\subsubsection{Current Challenges}
Interviews revealed the following two challenges in presenting network data live:

\noindent \textbf{\circled{C1} Limited Interactions: }
\rc{Presenting network data interactively, rather than simply navigating slides, poses challenges. 
These challenges arise from the complexity of network components and the need to design interactions to control each component while aligning the control with the narration.} As a result, existing presentations often only rely on static diagrams and pre-sequenced slides, which restrict direct interaction with the visualizations. While physical boards with stickers that can be progressively removed create a visually engaging effect, \ac{A2 remarked that \textit{``It takes a long time to ask the art department to set things up, and it's impossible to make adjustments mid-presentation.''}}
\rc{This lack of direct interaction makes it harder for presenters to guide the audience's focus through the interactions effectively. This limitation is critical since node-link diagrams lack a clear reading order.}

\noindent \textbf{\circled{C2} Interaction Disrupting Narrative Flow: }
A1, A3, and A5 noted challenges with \ac{multi-touch} display methods during live presentations.
\rc{Common issues include locating menus on large screens---particularly when presenting from the side---and the multiple steps required to switch presentation modes.} 
\ac{A3 explained, \textit{``With our touch-display setup, we have to click a menu at the screen's edge to select visuals, which takes time and disrupts the flow.''}}
These interactions require prior practice \ac{and remain prone to errors, even after practice.} 
\ac{Errors further interrupt presentations, requiring mid-presentation corrections that disrupt the narrative flow and detract from audience engagement.}

\subsection{Design Requirements}
Drawing from the current practices, identified challenges, and expert insights from the interviews, we defined the design requirements for a prototype supporting synchronous network data storytelling using physical objects as the interaction medium.

\noindent \textbf{\circled{R1} Flexible Interactions with Network Components:} The system should allow presenters to manipulate network components directly rather than relying on transitions between predefined states (e.g., slideshows). 
\rc{Direct manipulation of} network components is essential for effectively guiding the audience's attention and providing relevant context, especially given the lack of apparent reading orders in node-link diagrams. 
Allowing \ac{this flexibility supports improvisational storytelling, as emphasized in prior research~\cite{Chironomia, FromJamtoRecital}, and enables presenters to revisit the content, skip components, or dynamically adjust the layout. }
Such \rc{adaptability enriches} the audience's interactive and personalized experience while adapting to time constraints and other presentation conditions. These needs, identified by all participants, directly address the challenges outlined in \circled{C1}.

\noindent \textbf{\circled{R2} Simple and Intuitive Interactions: }The system should facilitate intuitive interactions with low cognitive loads, allowing presenters to focus on storytelling\ac{, aligning with prior studies' considerations~\cite{LeeSketchStory}.}
Simplifying interactions reduces the likelihood of mistakes and anxiety, ensuring ease of use during presentations. This approach makes the system accessible to a broader range of users with less need for prior knowledge. Interactions should be spontaneous and seamless to maintain the presentation's flow. Additionally, the system should avoid strict adherence to predefined body movements, which necessitate extensive rehearsals and increase the risk of errors\ac{, as discussed by Cao\etal~\cite{Elastica}}. These requirements, mentioned by three participants, address the challenges identified in \circled{C2}. 

\noindent \textbf{\circled{R3} Natural and Engaging Body Movements: }
\rc{The system should ensure that presenters' movements appear natural and engaging from the audience's perspective.} 
\ac{Participants noted that gestures should function not only as interaction triggers but also as enhancements to presentation quality. Awkward gestures can distract the audience from the content. A1 observed, \textit{``Using a hand gesture to advance slides might seem magical or feel out of place to an audience unfamiliar with the system's logic, drawing attention away from the content and end up being distracting.''} }
\rc{Participants suggested that interactions should fulfill necessary storytelling functions while remaining appealing to the audience. A5 emphasized that gradually revealing information by peeling away stickers guides the audience's focus and advances the storyline effectively. Similarly, A3 highlighted that tracing a trend in a bar chart with a pointer, while a line highlights the trend, can also effectively direct attention and advance the story. These insights, emphasized by all participants, highlight the importance of smooth, natural, and engaging interactions from the audience's perspective, addressing the challenges outlined in \circled{C2}.}

\noindent \textbf{\circled{R4} Manageable Physical Objects:} 
\rc{In discussions with participants, we identified key characteristics of physical objects as an interaction medium in live presentations.}
The system should limit \rc{both the number and types of objects} to \rc{reduce the learning curve and simplify handling}. The objects should be small enough to hold four or five in one hand, \rc{easy to grasp, and preferably thick and compact rather than card-like}. Additionally,  \rc{reusability is also essential to lower costs and reduce environmental impact.}
These factors contribute to the efficient, manageable, and sustainable use of physical objects in presentations.


\section{\ac{Soliciting Interactions}}
\label{section4}
\rc{Our formative study highlighted the importance of interactions with network visualizations and identified key design requirements for effective interaction.} 
Physical objects emerged as a promising approach, enabling presenters \rc{to manage network components easily and effectively.} 
Building on the insights from this study, 
\rc{we conducted workshops to investigate concrete mappings between \textbf{user actions} and \textbf{interaction commands}. 
\ac{Here, we define user actions as controls on physical objects and gestures (e.g., moving a physical object) and interaction commands as manipulations on network visualizations (e.g., repositioning a network node).}
Our goal is to define a comprehensive design space that captures the diverse possibilities of these mappings.}

\subsection{Interactions for Network Storytelling}
We first aimed to identify interaction commands (e.g., show a node or hide a \rc{link}) applicable to node-link diagrams in synchronous network-based data storytelling. 
\ac{We focus our research on the communication phase within the four stages of data storytelling proposed by Li\etal~\cite{dataStorySurveyHaotian}: analysis, planning, implementation, and communication.}
While we recognize that the network interaction commands required at each stage are not distinctly different, \fc{their purposes and emphasis differ.} 
\fc{In data analysis, commands such as annotating nodes support exploration by helping users track exploration and generate insights. In contrast, during presentations, the same commands can serve to emphasize key nodes and reinforce the narrative. Similarly, repositioning nodes is essential for untangling complex networks to understand their topologies, but in presentations, it is primarily used to guide audience attention and clarify relationships. 
To ensure our interaction commands align with real-time storytelling needs,
We compiled a set of fundamental interaction commands deemed most critical to enhancing the communication phase, while this list is not exhaustive.}
\rc{To this end, we first reviewed existing literature on network interaction tasks~\cite{orko, taskTaxonomyGraphBongshin, multitouchGraph}.} \ac{We then refined this list by examining existing network data storytelling authoring tools~\cite{bashDynamicGraphComic, kimDataToon, ExpressiveAuthoringNodeLink}, examples from various domains~\cite{NarratingNetworks}, and network story examples shared from news anchors (see~\cref{tab:experts_list}). The final set of interaction commands was specifically tailored for real-time interactions in network data storytelling.}


\subsection{Selection of Physical Objects}
\begin{figure}
    \centering
    \includegraphics[width=1\linewidth]{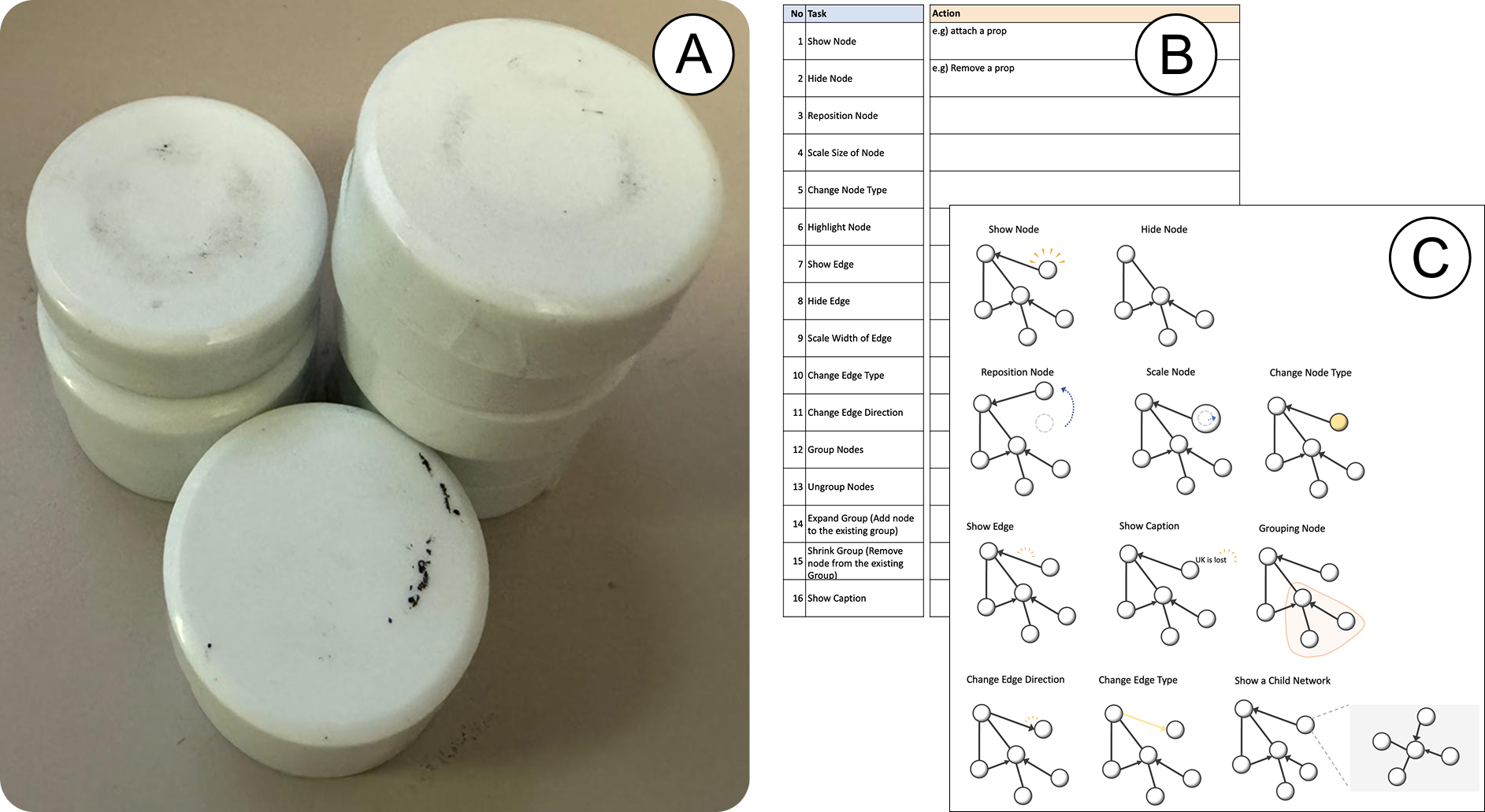}
    \caption{Workshop:
     Participants used double-sided magnets (A) to explore user actions for visualization commands and recorded their findings in worksheets (B). They referred to the illustration of visualization commands as needed (C). The ideas were then shared and discussed among the participants.
    }
    \label{fig:workshop}
    \Description{Figure 3: This figure includes a photograph of participants engaged in a workshop. It also shows materials used during the session, including worksheets for recording ideas, a handout with illustrations of visualization commands, and a picture of the double-sided magnets used for interactions.}
\end{figure}

Various physical objects can serve as interaction mediums for visualizations~\cite{bricks, PaperVis, DataCube, sphere}. Establishing a common reference is essential for grounding our discussion and exploring specific physical actions. We chose round double-sided 4×4cm magnets in our study (\cref{fig:workshop}-a). \rc{These magnets meet the requirements for ease of handling and cost-efficiency identified by the news anchors. Moreover, they offer versatile affordances, such as stacking and flipping.}
Magnets are widely used to represent networks in contexts like classrooms and sports strategy. 
\rc{While their physical properties afford unique interactions, many of these, such as stacking, flipping, and rotating, can be generalized to other objects with similar characteristics. This suggests our findings could extend to a broader range of physical objects.}

\subsection{Ideation Workshop}

\subsubsection{Participants}
In our workshop, 14 participants, comprising students and professors specializing in Visualization (VIS) and Human-Computer Interaction (HCI), took part. \ac{HCI/VIS experts were selected for their interaction design expertise, following methodologies from similar studies~\cite{DataCube, PaperVis}.} Participants, aged 23 to 38, included ten males and four females, as disclosed. We conducted five sessions with two to three participants each.

\subsubsection{Materials}
We provided participants with the 
set of interaction commands (e.g., \textit{Show/Hide Node}) and illustrations for each command to aid comprehension (\cref{fig:workshop}-b,c). These \rc{materials} were accessible to participants throughout the workshop. Additionally, participants received twelve 4×4 cm double-sided magnets (\cref{fig:workshop}-a) and were encouraged to experiment with them to explore potential user actions for \rc{the given interaction commands}.
In our preliminary workshop, we observed that participants occasionally confused interaction commands (e.g., \textit{Show Node}) with user actions (e.g., \textit{Attach a magnet}). To \rc{address this}, we introduced a few example mappings (see supplemental materials). This demonstration also served as a priming technique \cite{primingTech}, a method commonly used in previous studies \cite{DataCube, PaperVis}.


\subsubsection{Procedure}
Each 70-minute session was divided into three phases: briefing, individual brainstorming, and group discussion.
In the briefing, participants first completed a consent form and demographic questionnaire. We then provided a 10-minute overview of the research background and workshop objectives, emphasizing interaction in synchronous data storytelling. We explained each interaction command and demonstrated sample mappings, making it clear that these examples were only starting points and encouraging participants to think creatively beyond them.
During individual brainstorming (25–30 min), participants were tasked with developing ways to perform each interaction command using the provided magnets and recording their ideas in the worksheets. Participants were encouraged to disregard technical limitations.
In the group discussion (20–30 min), participants shared their ideas by demonstrating them with the magnets and discussed possible extensions. We encouraged them to explain the reasoning behind each interaction. The entire idea-sharing process was recorded and documented in the instructor's notes for further analysis.

\subsubsection{Data Analysis}
One author organized the workshop outcomes, identifying a total of 130 unique commands and their associated actions. \rc{Actions were initially coded using a reference set derived from prior studies~\cite{DataCube, PaperVis, TangibleTouch}}. \ac{This reference set included actions: Attach~\cite{DataCube}, Tap~\cite{DataCube, TangibleTouch}, Draw~\cite{DataCube, TangibleTouch}, Pinch~\cite{DataCube, TangibleTouch}, Stack~\cite{DataCube}, Collide~\cite{DataCube}, Point~\cite{PaperVis}, Bring Closer~\cite{DataCube}, Cover~\cite{DataCube, PaperVis}, Flip~\cite{PaperVis}, Rotate~\cite{DataCube, PaperVis}, and Relocate~\cite{DataCube, TangibleTouch, PaperVis}.} \rc{For actions beyond this initial set, an open-coding approach was applied, with codes adapted to reflect better actions involving magnets (e.g., Relocate was redefined as Slide). Coding was refined iteratively through discussions among three additional authors until consensus was achieved.}

\subsection{\ac{Design Space}}
\begin{figure*}[p]
    \centering
    \includegraphics[width=0.985\linewidth]{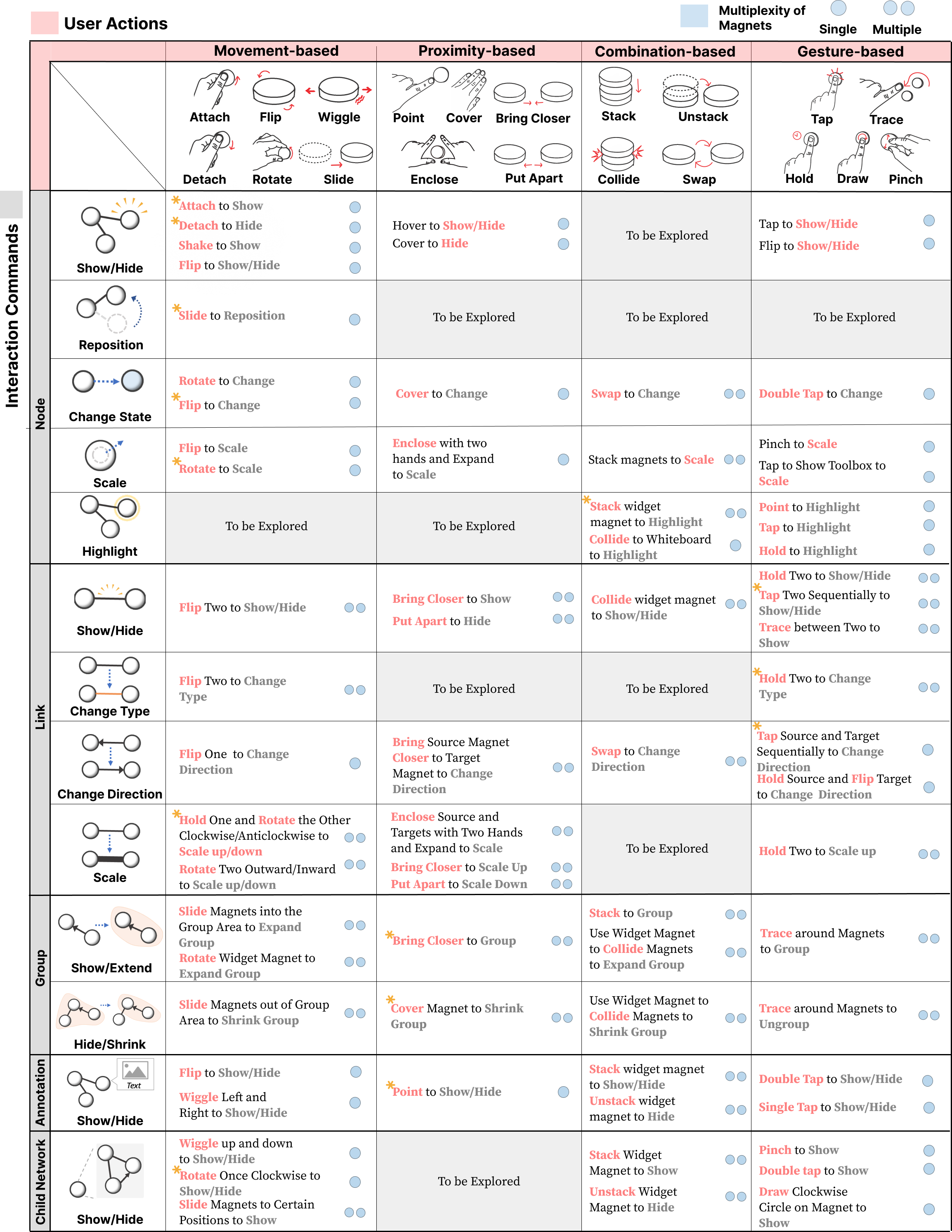}
    \caption{The design space maps user actions with double-sided magnets to interaction commands for network visualizations. The horizontal axis represents user actions, the vertical axis shows interaction commands, and each cell displays mappings. Icons in the cells indicate magnet multiplexity, as illustrated in the top-right legend. \ac{Mappings implemented in our prototype are marked with an asterisk.}
    }
    \Description{
    Figure 4: This figure presents a design space that maps user actions with double-sided magnets to network visualization commands. The horizontal axis lists user actions, and the vertical axis lists \rc{interaction commands for network visualizations}. Each cell in the grid indicates the identified mappings, with icons showing how many magnets are used for each interaction.
    }
    \label{fig:designspace}
\end{figure*}

Based on our analysis of the workshop results, we summarized a design space that maps user actions to interaction commands for network visualizations (\cref{fig:designspace}). This design space is structured around three dimensions: 1) Interaction Command, 2) Primary Modality of User Actions, and 3) Multiplexity of Physical Objects.

\subsubsection{Dimension 1: Interaction Command} 
This dimension defines interaction commands for manipulating network visualizations during live presentations, categorized by the network components they target (see~\cref{fig:designspace} row headers). \rc{These commands enable the manipulation of visualizations in real-time, supporting interactive and improvised storytelling.}

Presenters can show or hide nodes as the narrative unfolds, such as when introducing new characters or removing them (\textit{i.e., Show/Hide Node}). Nodes can also be repositioned to reduce clutter or better represent relationships (\textit{i.e., Reposition Node}). 
Adjusting node size can signify changes in attributes (\textit{i.e., Scale Node}) while altering a node's visuals (e.g., shape or color) can indicate changes in its state (\textit{i.e., Change Node State}).
\rc{Links} require similar dynamic manipulation, such as toggling their visibility (\textit{i.e., Show/Hide \rc{Link}}), changing their state (\textit{i.e., Change \rc{Link} Type}), \rc{or adjusting their width (\textit{i.e., Scale \rc{Link}}) to match the narrative.}
The direction of \rc{links} is critical in illustrating relationships between nodes and can also be modified \rc{during presentations} (\textit{i.e., Change \rc{Link} Direction}).
The dynamic grouping of nodes is another essential aspect of network data storytelling \ac{(\textit{i.e., Show/Extend Group})}. For example, nodes associated with a particular theme, such as political parties or product categories, are often grouped with visual boundaries~\cite{kimDataToon}. 
Nodes can be added to or removed from these groups as the narrative progresses \ac{(\textit{i.e., Hide/Shrink Group})}.
Annotations provide contextual information, helping the audience understand the story~\textit{i.e., Show/Hide Annotation}). Similarly, child networks, representing lower-layer networks within nodes, can be revealed (\textit{i.e., Show/Hide Child Network}). In an international relations network, for example, a presenter might zoom into a country to explore its local government relationships, as suggested by news anchors during our interviews.

\subsubsection{Dimension 2:  Primary Modality of User Actions} 

This dimension encompasses various actions that users can perform (see~\cref{fig:designspace} column headers) with the magnets.
The actions are further classified into four categories:
\textbf{Movement-based Interactions} involve direct physical manipulation of the magnet, where users alter the magnet's position, orientation, or motion. Actions in this category include \textit{Attach}, \textit{Detach}, \ac{\textit{Slide}}, \textit{Rotate}, \textit{Wiggle}, and \textit{Flip} (\cref{tab:action_M}-A). 
\textbf{Proximity-based Interactions} depend on the spatial relationships between magnets or between magnets and hands, focusing on their relative distances. Examples include \textit{Bring Closer}, \textit{Pull Apart}, \textit{Cover}, \textit{Point} and \textit{Enclose} (\cref{tab:action_M}-B). 
\textbf{Combination-based Interactions} involve the simultaneous or sequential use of multiple magnets, such as \textit{Stack}, \textit{Unstack}, \textit{Collide}, and \textit{Swap} (\cref{tab:action_M}-C). 
\textbf{Gesture-based Interactions} refer to gestures performed by hands in relation to the physical objects, taking into account their spatial relationships. Examples include \textit{Tap}, \textit{Draw}, \textit{Trace}, \textit{Pinch}, and \textit{Hold} (\cref{tab:action_M}-D).
\ac{By categorizing user actions in this way, we focus on single, discrete actions that users can perform, which serves as the foundation for mapping to interaction commands.}

\begin{table*}
\centering
\caption{User actions by primary modality: We categorized user actions into four key dimensions: Movement-based Interactions, Proximity-based Interactions, Combination-based Interactions, and Gestural Interactions.}

\begin{tabular}{p{0.48\linewidth}p{0.48\linewidth}}
 \Xhline{3\arrayrulewidth}
 
\vspace{0.01\linewidth}
\noindent\textbf{(A) Movement-based Interactions}: 
\vspace{0.04\linewidth}

\begin{tabular}{m{0.05\textwidth}m{0.4\textwidth}}  
    \includegraphics[width=0.05\textwidth, height=0.05\textwidth]{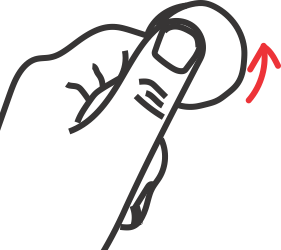} &
    \textbf{Attach}:  Placing a magnet onto the interaction space. \\
    \includegraphics[width=0.05\textwidth, height=0.05\textwidth]{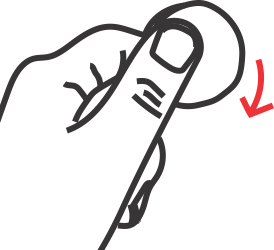} &
    \textbf{Detach}: The opposite of attaching; removing the magnet from the interaction space. \\
    \includegraphics[width=0.05\textwidth, height=0.05\textwidth]{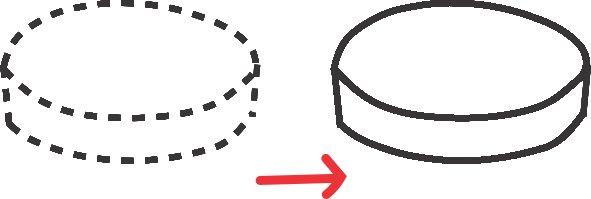} &
    \rc{\textbf{Slide}}: Moving the magnet from one location to another without detaching it. \\
    \includegraphics[width=0.06\textwidth, height=0.04\textwidth]{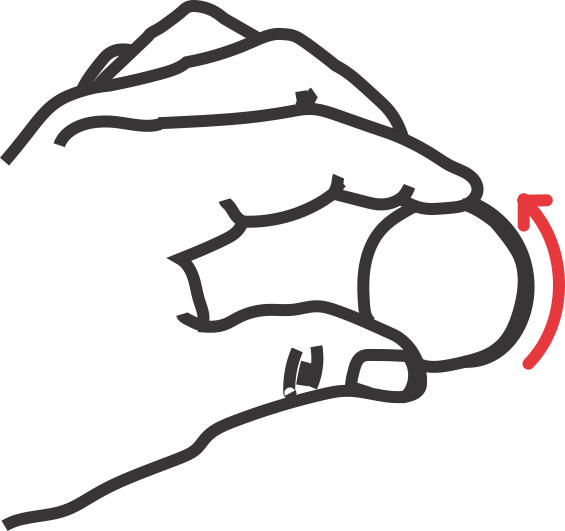} &
    \textbf{Rotate}: Turning the magnet around its axis, changing its orientation without altering its location. \\
    \includegraphics[width=0.06\textwidth, height=0.05\textwidth]{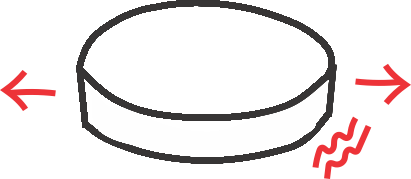} &
    \textbf{Wiggle}: Slightly shifting the magnet vertically or horizontally while it stays attached to the interaction surface. \\
    \includegraphics[width=0.05\textwidth, height=0.04\textwidth]{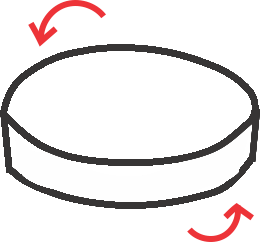} &
    \textbf{Flip}: Inverting the magnet by turning it over to reveal the opposite surface. \\
\end{tabular}

&

\vspace{0.01\linewidth}
\noindent \textbf{(B) Proximity-based Interactions}: 
\vspace{0.04\linewidth}

\begin{tabular}{m{0.05\textwidth}m{0.4\textwidth}}
    \includegraphics[width=0.06\textwidth, height=0.04\textwidth]{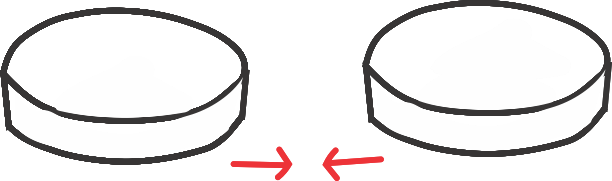} &
    \textbf{Bring Closer}: Moving multiple magnets toward each other without making contact. \\
    \includegraphics[width=0.06\textwidth, height=0.04\textwidth]{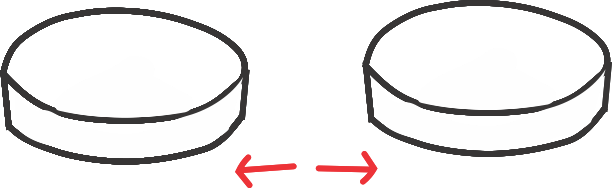} &
    \textbf{Pull Apart}: Moving multiple magnets away from each other. \\
    \includegraphics[width=0.05\textwidth, height=0.05\textwidth]{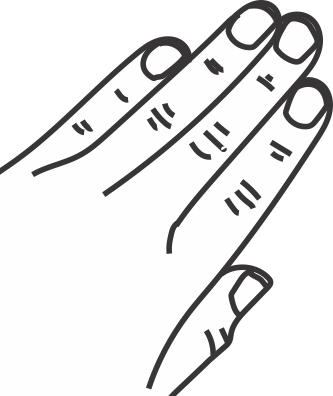} &
    \textbf{Cover}: Positioning a magnet or hand over the magnet without making direct contact to obscure it from view. \\
    \includegraphics[width=0.05\textwidth, height=0.05\textwidth]{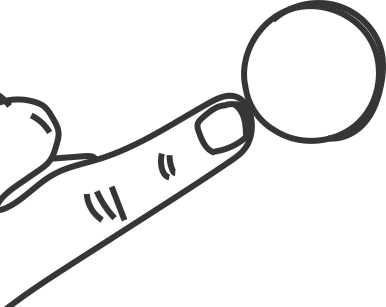} &
    \textbf{Point}: Pointing near the magnet with a finger without making direct contact. \\
     \includegraphics[width=0.07\textwidth, height=0.05\textwidth]{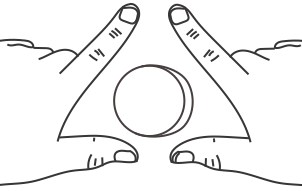} &
    \textbf{Enclose}: Using hands to form a boundary around the proximity of magnets. \\
\end{tabular}
\vspace{0.01\linewidth}
\\  \hline
\vspace{0.01\linewidth}
\noindent \textbf{(C) Combination-based Interactions}: 
\vspace{0.04\linewidth}

\begin{tabular}{m{0.05\textwidth}m{0.4\textwidth}}
    \includegraphics[width=0.05\textwidth, height=0.05\textwidth]{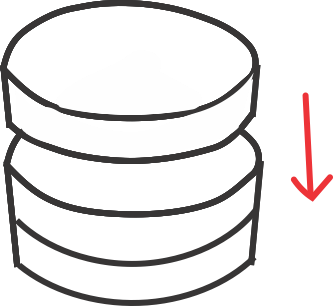} &
    \textbf{Stack}: Placing one magnet on top of another. \\
    \includegraphics[width=0.05\textwidth, height=0.05\textwidth]{figs/hand/stack.png} &
    \textbf{Unstack}: The opposite of stacking; removing a magnet from the stack. \\
    \includegraphics[width=0.05\textwidth, height=0.05\textwidth]{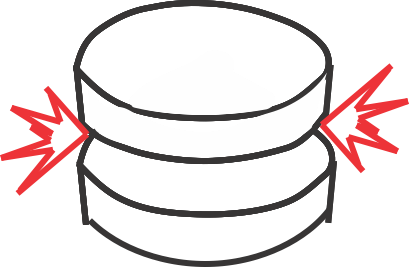} &
    \textbf{Collide}: Deliberately causing two magnets to impact each other. \\
    \includegraphics[width=0.06\textwidth, height=0.04\textwidth]{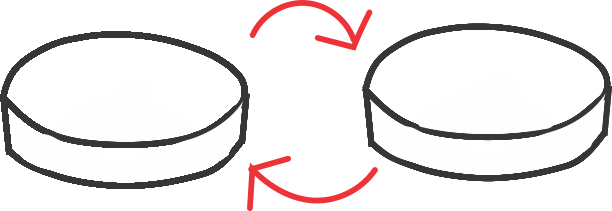} &
    \textbf{Swap}: Exchanging the positions of two magnets. \\
\end{tabular}

& 

\vspace{0.01\linewidth}
\noindent \textbf{(D) Gestural Interactions}: 
\vspace{0.04\linewidth}

\begin{tabular}{m{0.05\textwidth}m{0.4\textwidth}}
    \includegraphics[width=0.05\textwidth, height=0.06\textwidth]{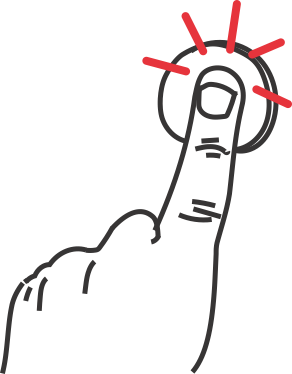} &
    \textbf{Tap}: Quickly touching and releasing the surface of the magnet with a finger. \\
    \includegraphics[width=0.05\textwidth, height=0.06\textwidth]{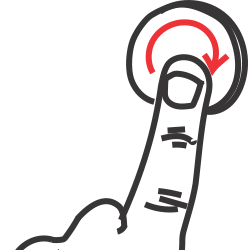} &
    \textbf{Draw}: Using a finger to trace a specific path or trajectory on the surface of the magnet. \\
    \includegraphics[width=0.06\textwidth, height=0.05\textwidth]{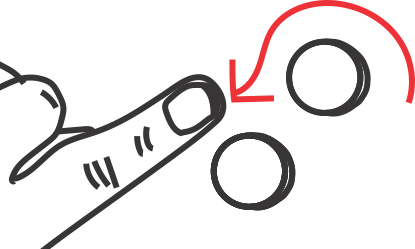} &
    \textbf{Trace}: Using a finger to follow the contour or outline of an object with one or more magnets. \\
    \includegraphics[width=0.05\textwidth, height=0.05\textwidth]{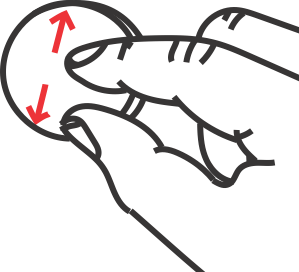} &
    \textbf{Pinch}: Using the thumb and index finger to establish two contact points on the magnet's surface, then bringing those points closer together or pulling them apart. \\
    \includegraphics[width=0.05\textwidth, height=0.06\textwidth]{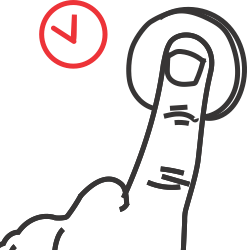} &
    \textbf{Hold}: Maintaining continuous contact with the magnet's surface using a finger for a sustained period. \\
\end{tabular}
\vspace{0.01\linewidth}

\\  \Xhline{3\arrayrulewidth}
\end{tabular}
\label{tab:action_M}
\end{table*}

\subsubsection{Dimension 3: Multiplexity of Physical Objects}
\rc{This dimension classifies actions as involving a \textbf{Single Object} or \textbf{Multiple Objects}, with the latter further divided into \underline{\textit{Simultaneous}} and \underline{\textit{Sequential}} Actions (see~\cref{fig:designspace} icons).} \ac{It clarifies the magnet multiplexity and the temporal coordination required when using multiple magnets}

\noindent \textbf{Single Object:} 
These actions involve a single magnet, typically manipulated with one hand, such as moving, flipping, or rotating a magnet. They are straightforward and do not require coordination with other magnets.

\noindent \textbf{Multiple Objects:} 
These actions involve multiple magnets. While these actions introduce complexity, they also enhance expressiveness. Their outcome often depends on the temporal coordination of multiple magnets and typically targets links or groups.

\noindent\underline{\textit{Simultaneous Actions}}: 
\rc{These actions involve multiple magnets simultaneously or have outcomes unaffected by order.} \ac{For example, flipping two magnets simultaneously to change a link type or bringing two magnets closer to form a group.} \ac{While they do not require temporal coordination,} they often demand the use of both hands, adding a layer of physical coordination.

\noindent\underline{\textit{Sequential Actions}}: 
These actions require performing operations on multiple magnets in a specific sequence. \ac{For instance, tapping magnets sequentially can define link directionality.} 
The outcome depends on the action order, requiring users to follow a predefined sequence to achieve the desired effect.
\ac{Sequential actions offer greater expressiveness than simultaneous ones, as they allow for specifying directions.}

\begin{figure*}[t!]
    \centering
    \includegraphics[width=0.85\linewidth]{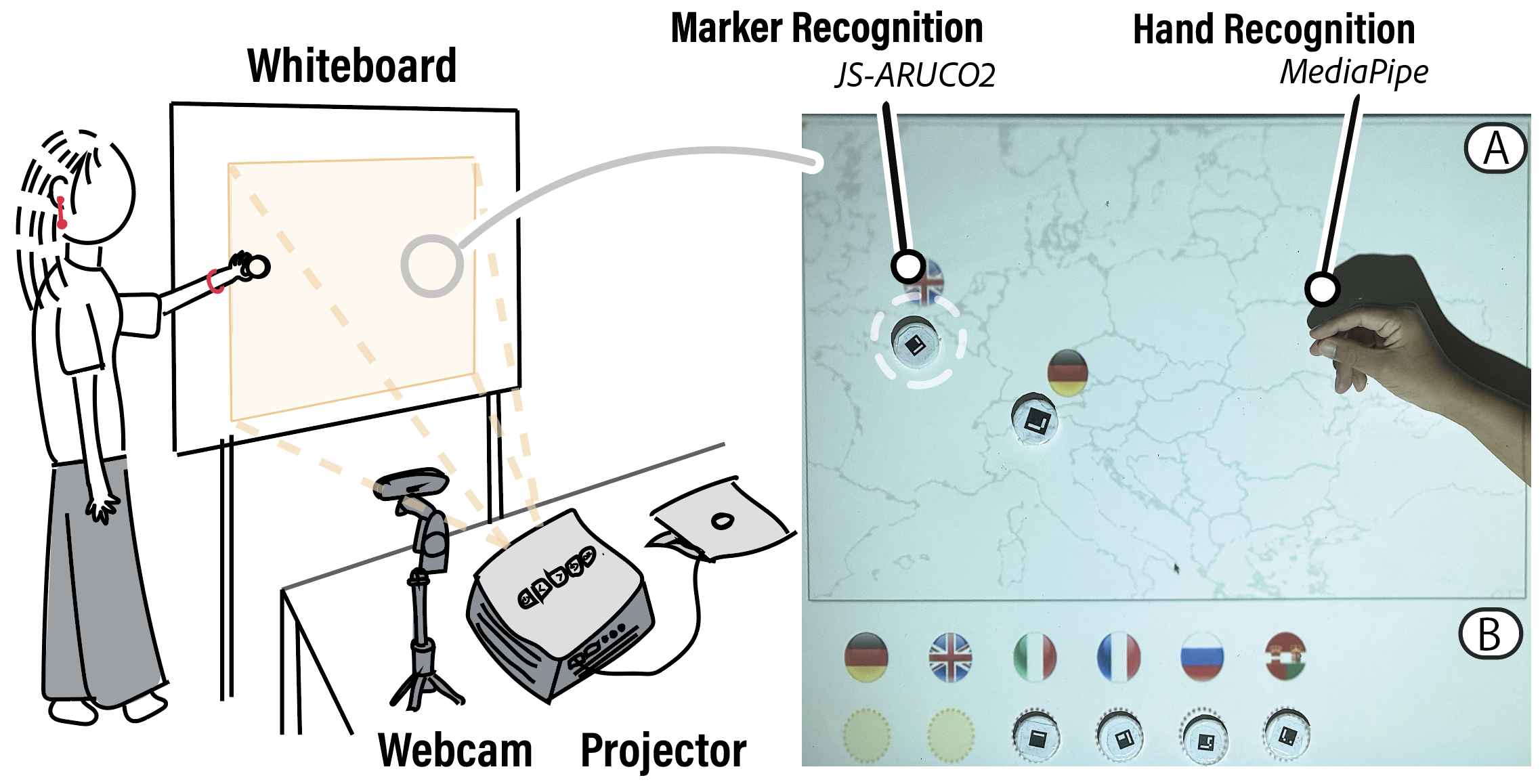}
    \caption{System Overview: TangibleNet recognizes the presenter's interactions with magnets and hand gestures on the whiteboard via a webcam and projects the resulting visualizations onto the whiteboard using a projector. The projected user interface is divided into two areas: (A) storyboard area and (B) registration area.}
    \label{fig:systemsetting}
    \Description{
    Figure 5: This figure provides both an illustration and a photograph of the TangibleNet system setup. In the illustration, the presenter stands in front of a whiteboard while a webcam on a nearby table tracks the movements of the presenter and the physical markers on the board. A projector, also on the table, projects virtual objects onto the whiteboard. The user interface projected onto the board is divided into two sections: the storyboard area and the registration area. Further details about these sections can be found in Sections 5 and 6.
    }
\end{figure*}

\section{Prototyping and Refining Interaction Design}
\label{codesign}
We iteratively developed the prototype based on design requirements from the formative study and design space from workshops. 
From the broad design space, we mapped visualization commands to user actions, ensuring they met design requirements and incorporated critical considerations for effective communication. Throughout this iterative process, we consistently received feedback from the two news anchors \ac{(A1 and A5)} who participated in the initial study, helping to refine our design.

\noindent \textbf{Interaction from the Presenter's Perspective: }Presenters need simple, low-cognitive-load interactions to maintain a smooth presentation flow \circled{R2}. Complex, multi-step interactions increase operational difficulty and disrupt the delivery. Feedback from news anchors highlighted that interactions requiring presenters to navigate multiple steps, such as activating a toolbox, were undesirable for maintaining a smooth presentation. Similarly, assigning the same user action to different interaction commands through mode selection complicated the process. Consequently, we aimed to avoid including mode choices and ensured that each interaction command was clearly assigned to distinct user actions. To effectively guide audience attention \circled{R1}, we aligned interaction commands with the presenter's deictic gestures. For example, we implemented a gesture where pointing at a magnet triggers the display of an annotation, \rc{directing the audience's} focus to the relevant area. This design is based on practices used by news anchors, who point at visuals while cueing studio staff to display corresponding content.

\noindent \textbf{Interaction from the Audience's Perspective: }
The news anchors raised concerns about several mappings that appeared unnatural from the audience's perspective \circled{R3}. For instance, we initially implemented the \textit{Show/Hide \rc{Link}} feature by bringing two nodes closer together, which seemed intuitive for presenters as a toggle action \circled{R2}. 
\rc{However, the anchors found this misleading to the audience -- bringing nodes closer suggests a stronger relationship, not disappearing or weakening \rc{link}. This mismatch may evoke confusion, as the presenter's gestures did not produce the expected visual change for the audience.}
Another example of problematic mappings was using a pinching gesture on a magnet to reveal a child network. 
Pinching was too subtle for the audience, making transitions feel abrupt. 
This feedback highlighted the need for interactions that align with both the presenter's intent and the audience's expectations.  

\noindent \textbf{Dynamic Mapping Magnets and Nodes: }
A network data story typically involves multiple nodes, making it challenging to link magnets to specific nodes. A simple solution is to assign magnets to nodes based on a predefined sequence, but this limits flexibility for improvised storytelling \circled{R1}. Alternatively, a one-to-one mapping between magnets and nodes can be established before the presentation. This method allows for a more flexible order in which nodes are shown, but it requires users to memorize the assignments.
Adding distinguishable features, such as stickers, to the magnets reduces reusability and increases preparation time, contradicting the design requirement \circled{R4}.
To address these issues, we designated a registration area where presenters place magnets to assign them to nodes dynamically. 
This approach reduces the presenters' cognitive load of memorizing the assignments while allowing flexible storytelling. 

\noindent \textbf{Data Registration before Presentaton: }
We found that assigning a magnet to every node overwhelmed users and disrupted the presentation flow, as manually displaying all nodes was time-consuming. To address this, we pre-registered the network data, following practices from previous studies \cite{LeeSketchStory, Chironomia}. Discussions with news anchors revealed that network data stories often include nodes of varying importance, with some serving as background information for key nodes. We categorized the nodes into primary and secondary, where magnets are only assigned to primary nodes, while secondary nodes appear automatically alongside primary nodes. Although this limited the direct manipulation of secondary nodes, it reduced the number of magnets needed, streamlining the presentation without compromising the story's flow \circled{R4}. 
Similarly, \rc{links} can also be considered secondary and follow the appearance of the primary nodes to enhance narrative continuity and minimize tedious interactions.

\noindent \textbf{Accessible Physical Interactions: }
To enhance accessibility, we chose a Computer Vision (CV)--based approach over electronic circuits.
Using ArUco markers\footnote{\url{https://github.com/fdcl-gwu/aruco-markers}}, the system captures each marker’s identity, position, and orientation while recognizing hand gestures (e.g., touch and hold).
This approach allows easy deployment in diverse environments \circled{R4}, such as classrooms and offices, without requiring electronics expertise~\cite{ARUCODesignSpace} or expensive wall-sized touch displays (e.g., smart boards). However, CV--based techniques are constrained by occlusions and glitches to subtle changes. 
This choice limits the inclusion of certain physical actions, such as wiggling and colliding), which were not included in the prototype.
\setlength{\intextsep}{0pt} 

\section{TangibleNet}
Building on the prototyping process, we present our prototype, TangibleNet. TangibleNet consists of a webcam, double-sided magnets with ArUco markers, a projector, and a whiteboard (\cref{fig:systemsetting}). Users interact with visualizations by manipulating the magnets on the whiteboard and performing user actions. The webcam captures these actions, which are processed to manipulate network visualizations. The visualizations are projected onto the whiteboard, aligned with the magnets' positions.

\subsection{User Interface}

The user interface consists of two areas: registration and storyboard (\cref{fig:systemsetting}). The registration area displays primary nodes from the pre-loaded network data, along with highlighted spots for placing magnets (\cref{fig:systemsetting}-b). 
Placing a magnet dynamically registers it to the corresponding node. 
The storyboard area shows the network visualization, allowing presenters to manipulate it using magnets and hand gestures (\cref{fig:systemsetting}-a).

\subsection{Implementation Details}
TangibleNet is a browser-based application developed in JavaScript, with a backend powered by Node.js\footnote{\url{https://nodejs.org/}}. The system captures video streams from a webcam to detect the position and rotation of magnets using the OpenCV-based ArUco marker detection library, JS-ARUCO2\footnote{\url{https://github.com/damianofalcioni/js-aruco2/}}. Additionally, hand gestures are recognized using MediaPipe\footnote{\url{https://ai.google.dev/edge/mediapipe/}}, 
which tracks the presenter’s index finger to determine when it touches a magnet and measures the duration of contact by monitoring the continuity of this interaction. These interactions are reflected in the network visualization rendered by d3.js\footnote{\url{https://d3js.org}}, which is projected onto a whiteboard. 
To prevent overlapping ArUco markers and nodes that could hinder marker recognition accuracy, the nodes are projected near the magnets rather than directly on top of them.

\subsection{Interaction on TangibleNet}
\begin{figure*}[t!]
    \centering
    \includegraphics[width=1\linewidth]{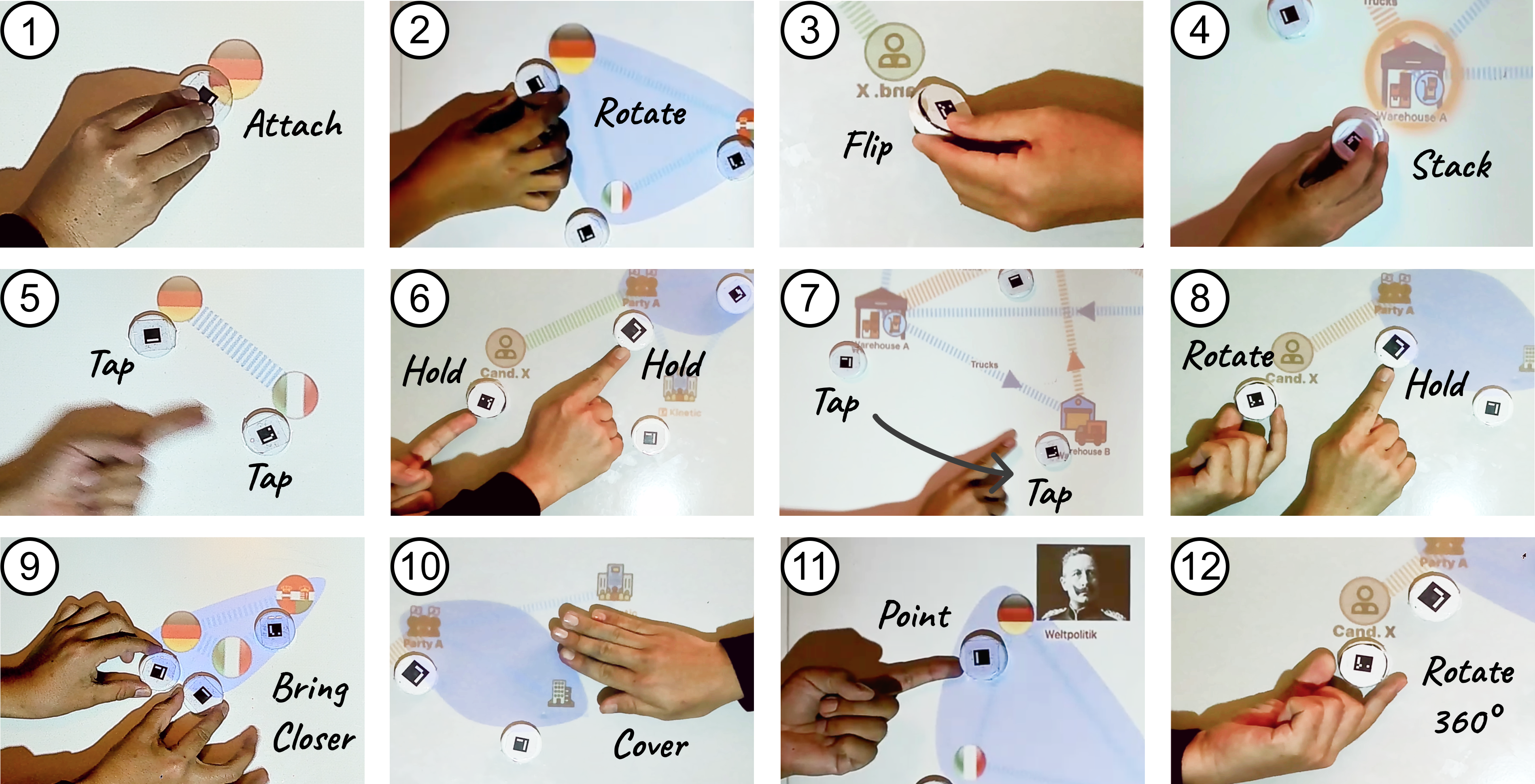}
\caption{\ac{Supported interactions include: (1) Attaching or detaching magnets to show or hide nodes; (2) Rotating a magnet to scale a node; (3) Flipping a magnet to change a node's state; (4) Stacking a magnet to highlight a node; (5) Tapping magnets sequentially to show or hide links; (6) Holding magnets simultaneously to change link types; (7) Tapping the source magnet first and then the target magnet to change link direction; (8) Holding one magnet and rotating the other to scale the link; (9) Bringing magnets closer to show or extend a group; (10) Covering a magnet to hide or shrink a group; (11) Pointing a magnet to display an annotation; (12) Rotating a magnet 360° to show or hide a child network.}}
    \label{fig:tangibleNet}
    \Description{
   Figure 6: This figure illustrates the various interaction commands supported by TangibleNet, including Show/Hide Node, Reposition Node, Rescale Node, Change Node State, Show/Hide \rc{Link}, Show/Hide Annotation, Show/Extend Group, Hide/Shrink Group, and Show/Hide Child Network. The methods for executing these commands are explained in Section 6.3.
    }
\end{figure*}

TangibleNet implements the interaction commands from our design space, mapping each to corresponding user actions (see \cref{fig:tangibleNet}). 
As noted in \cref{codesign}, CV constraints limited the range of user actions, making it difficult to implement all commands unambiguously due to recognition errors in hand gestures and marker detection. 
To address this, we developed multiple command sets containing interaction commands that are unambiguously mapped to user actions. Users can select their command set before their presentation. 

\subsection{\ac{Use Scenarios}}
\label{usecase}
We illustrate the applicability of our design space through various storytelling scenarios, showcasing TangibleNet in action. Supplementary videos provide visual demonstrations.

\subsubsection{\ac{Case 1: Shifting Alliances during World War I}}
\ac{Inspired by prior studies~\cite{bashDynamicGraphComic, kimDataToon, DGComics}, this case examines evolving alliances in World War I. Countries are represented as nodes, and their alliances and hostilities as links.}

\textbf{Use Scenario:} Alex, a middle school history teacher, uses TangibleNet to teach her students about the evolving alliances in World War I.
First, she picks up several magnets and places them in the registration area to register magnets as nodes representing countries. She then slides the magnets representing Germany and Austria to the storyboard area, triggering the \textbf{\textit{Show Node}} command (\cref{fig:tangibleNet}-1).
As she explains the Dual Alliance, she taps the Germany and Austria magnets sequentially, triggering the \textbf{\textit{Show Link}} command (\cref{fig:tangibleNet}-5) and guiding student focus.
She then explains how Italy, due to territorial disputes, aligned with Germany. She moves Italy's magnet closer to Germany's, triggering the \textbf{\textit{Show/Extend Group}} command to group the countries (\cref{fig:tangibleNet}-9). 
As tensions escalate, she describes Germany's military expansion. She rotates its magnet clockwise to trigger the \textbf{\textit{Scale Node}} command, scaling up its node to emphasize its rising power (\cref{fig:tangibleNet}-2). 
When a student asks about Germany's military strength, she holds her finger on German's magnet to trigger the \textbf{\textit{Show Annotation}} command to provide more details (\cref{fig:tangibleNet}-11).
This interactivity allows her to flexibly adapt the story to align with her students' interests. Continuing, she points at Serbia's magnet to display an annotation about the assassination of Archduke Franz Ferdinand, explaining its role in triggering the war. As links multiply and cause visual clutter, she slides magnets to trigger the \textbf{\textit{Reposition Nodes}} command, adjusting the layout for clarity.

\subsubsection{\ac{Case 2: Explaining Political Alliances During an Election}}
\ac{Inspired by discussions with news anchors and their real-world examples, this case examines shifting political relationships among politicians, parties, and corporations during an election campaign. Nodes represent these entities, while links depict endorsements, alliances, and funding connections.}

\textbf{Use Scenario:} Emma, a news anchor, prepares for a live broadcast on election dynamics in the upcoming election. 
She attaches the magnets representing Party A, Candidate X, and major corporations to the storyboard area, triggering the \textbf{\textit{Show Node}} command. To illustrate affiliations, she taps Candidate X and Party A sequentially, triggering the \textbf{\textit{Show Link}} command to display the link indicating that Candidate X belongs to Party A. She then moves Party A and a corporation closer, triggering the \textbf{\textit{Show/Extend Group}} command to indicate an endorsement or funding relationship.
As the campaign progresses, she highlights substantial funding from Party A to Candidate X. She holds Candidate X's magnet and rotates the corporation's magnet outward, activating the \textbf{\textit{Scale Link}} command to visually emphasize the strength of financial connections by thickening the link between Candidate X and the corporation (\cref{fig:tangibleNet}-8). 
To explore Candidate X's support network, she rotates Candidate X's magnet 360 degrees clockwise, triggering the \textbf{\textit{Show Child Network}} command (\cref{fig:tangibleNet}-12), revealing a child network showing relationships with lobbyists, interest groups, and grassroots organizations supporting the candidate. 
At a pivotal point in the campaign, she explains that a scandal involving Party A has caused several corporations to withdraw their political alliances with the party. 
She covers their magnets to trigger the \textbf{\textit{Shrink/Hide Group}} command, removing them from the group associated with Party A (\cref{fig:tangibleNet}-10).
Additionally, she emphasizes a change in Candidate X's political stance due to the scandal. 
She flips Candidate X's magnet to trigger the \textbf{\textit{Change Node State}} command (\cref{fig:tangibleNet}-3), which changes the node's visual representation to reflect a policy shift. Finally, she holds both the magnets of Candidate X and Party A simultaneously, triggering the \textbf{\textit{Change Link Type}} command (\cref{fig:tangibleNet}-6), modifying the link's color to signify their shift from allies to opponents visually.

\subsubsection{\ac{Case 3: Optimizing a Supply Chain Network}}
\ac{This case visualizes a supply chain network to identify bottlenecks and improve logistics. Nodes represent factories, distribution centers, and retailers, while links depict the flow of goods.}

\textbf{Use Scenario:} Linda, a logistics manager, presents strategies to improve supply chain efficiency. She begins by attaching magnets representing factories, distribution centers, and retail stores to the whiteboard, triggering the \textbf{\textit{Show Node}} command. Links representing existing transportation routes appear automatically.
Narrating a specific distribution center experiencing reduced capacity and acting as a bottleneck, Linda stacks a widget magnet on its node, activating the \textbf{\textit{Highlight Node}} command (\cref{fig:tangibleNet}-4). She then holds the bottlenecked center's magnet and rotates the affected retail store's magnet inward, triggering the \textbf{\textit{Scale Link}} command. The link between them becomes thinner, visually indicating the reduced flow of goods on that route.
To propose an alternative, Linda taps the magnets of other distribution centers sequentially, activating the \textbf{\textit{Change Link Direction}} command (\cref{fig:tangibleNet}-7). This adjustment redirects the flow of goods from the overloaded center to others, providing a visual representation of the proposed optimization.

\begin{figure*}[t!]
    \centering
    \includegraphics[width=1\linewidth]{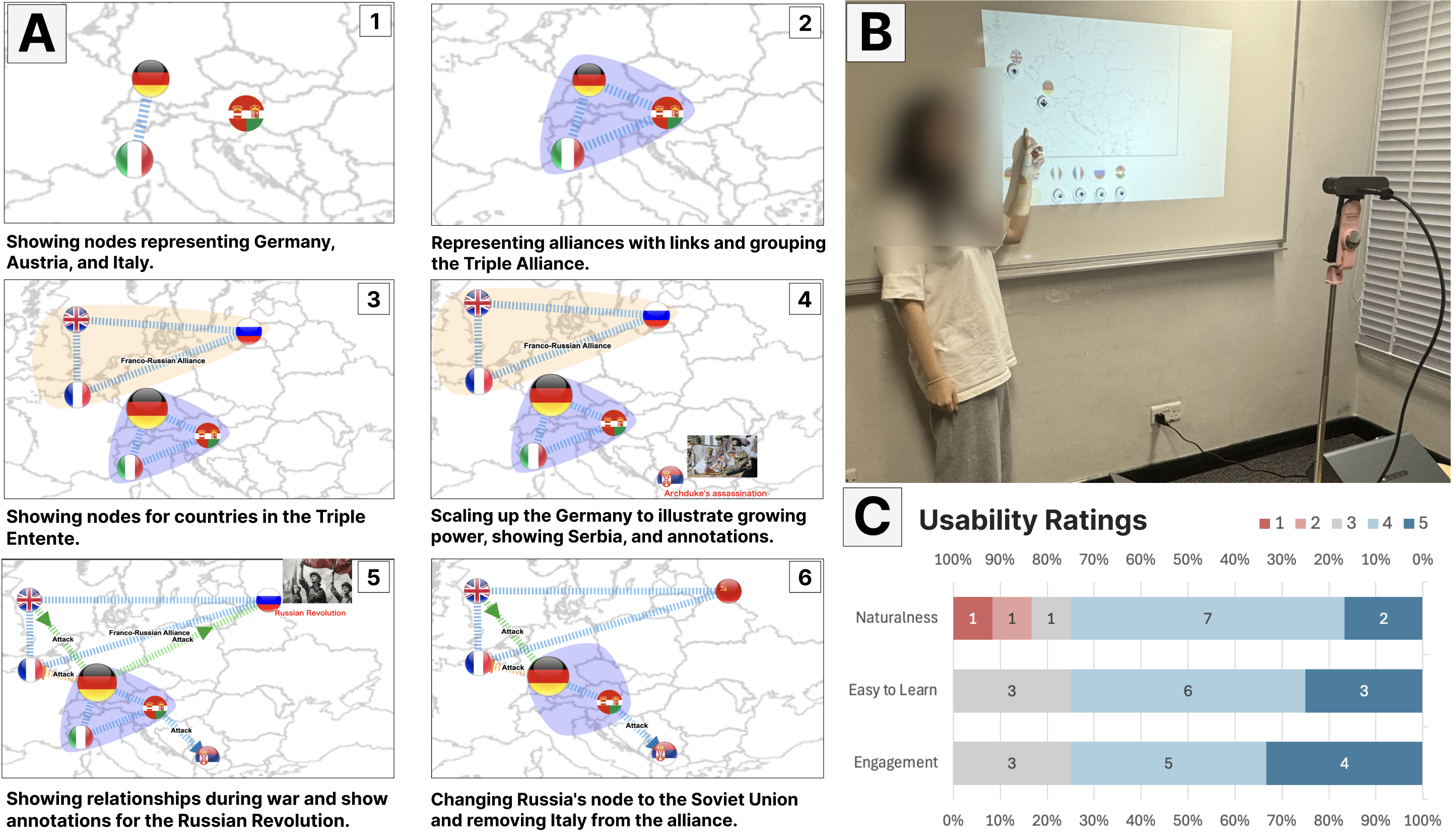}
    \caption{Evaluation setting: (A) Participants presenting the story using TangibleNet. (B) Selected snapshots of the target network data story, progressing sequentially from 1 to 6. (C) Ratings of TangibleNet on a 1–5 scale for the naturalness of the interactions, ease of learning, and overall engagement, with 1 being the least favorable and 5 the most favorable.}
    \label{fig:experiment}
    \Description{
   Figure 7: This figure shows a participant presenting with TangibleNet alongside a visual representation of how the network-based data story evolves. The story demonstrates shifting alliances leading up to and during World War I. In Scene 1, only Italy, Germany, and Austria are visible. In Scene 2, these countries form a group. Scene 3 introduces the UK, France, and Russia, which also form a group. Scene 4 introduces Serbia, along with an annotation depicting the assassination of Archduke Franz Ferdinand. In Scene 5, arrows appear between Germany and neighboring countries. In Scene 6, Russia transforms into the Soviet Union, and Italy exits the war.
    }
\end{figure*}

\section{Evaluation}
We conducted a user study to evaluate the presentation experience of network-based data storytelling with TangibleNet, gathering feedback on its utility, usability, and learnability. Additionally, we compared it to participants' familiar presentation environments and explored potential applications. While authoring support is important, this study does not address it.



\subsection{Participants}
Our study included 12 participants (P1-P12) recruited from the local community, comprising eight males, three females, and one non-binary individual. Participants held a range of occupations, including university students, public officers, software engineers, and television directors, with ages ranging from 20 to 34. Most participants had experience delivering live presentations that involved data, and all were familiar with node-link diagram representations. Six participants had experience using these diagrams in their presentations. Participants' familiarity with AR systems varied widely, from no prior knowledge to developing AR applications. 

\subsection{Apparatus}
We used a Logitech HD 1080p webcam with a 60fps frame rate. The projector used was an OWNKNEW model with 27,000 lumens and a resolution of 1920x1080p, projecting onto a whiteboard measuring approximately 70cm by 120cm. The system was run on a MacBook Pro (2021) with an Apple M1 Pro chip and 16GB of RAM. The magnets used in the setup had a diameter of 4cm and a thickness of 1.5cm.

\subsection{Tasks and Procedure}
Participants used TangibleNet to present a network data story (\cref{fig:experiment}-b). We chose the shifting alliances before and during World War I (\cref{fig:experiment}-a) as the target network story (see supplemental materials for the complete storyline). This scenario was considered appropriate because it encompasses a wide range of interaction commands and has been frequently studied in network data storytelling research~\cite{kimDataToon, DGComics,bashDynamicGraphComic}. 
The story incorporates commands: \textit{Show/Hide Node}, \textit{Reposition Node}, \textit{Scale Node}, \textit{Change Node State}, \textit{Show/Hide Link}, \textit{Show/Extend Group}, \textit{Hide/Shrink Group}, \textit{Show Annotation}, and \textit{Show/Hide Child Network}.

Each session began with participants completing a consent form and demographic questionnaire. They were informed they could withdraw at any time. Instructors then explained the experimental setup, explaining that participants would present in front of a whiteboard with projected visualizations. 
The instructors introduced the basic concepts of network data storytelling, including node-link diagrams and interaction commands. 
Participants were also given a handout that outlined the target network data story to help them familiarize themselves with the narrative (see supplemental materials). Following this, the instructors demonstrated TangibleNet, showcasing the interaction techniques available in the system. Participants were then asked to present the story using TangibleNet. During the presentation, the instructor posed questions (e.g., \textit{``Could you show that change again?''}) and requested unscripted manipulations to demonstrate TangibleNet's flexibility. We told participants they could ask questions about the interactions throughout their presentations if needed. 
Following the presentation, \ac{participants rated TangibleNet on a 1-5 scale based on the naturalness of the interactions, ease of learning, and overall engagement.}
These ratings informed a subsequent semi-structured interview, where participants elaborated on their experiences with the system.
\rc{They were also asked to compare TangibleNet with their own familiar presentation environments, providing insights into areas for improvement and potential application scenarios.} Each user study session lasted approximately 50 to 60 minutes, and participants received 6 USD as compensation.


\subsection{Observation and Feedback}
All participants completed their presentations using TangibleNet, interacting with the network visualization throughout. \ac{They provided subjective feedback, using their familiar workflows as an informal baseline where applicable.} Below, we discuss the insights gathered from the study.

\noindent \textbf{\rc{Easy-to-Learn and Straightforward Interactions:}} 
Most participants were able to interact with the network visualization immediately after a single demonstration by the instructor. \ac{This ease of use is reflected in the responses, where 9 out of 12 participants agreed that the system was easy to learn, and none disagreed (\cref{fig:experiment}-c).} 
\rc{Only P9 and P10, who had no prior experience with AR, required additional clarification.} This underscores TangibleNet's low learning curve.
P1 remarked, \textit{``I can't think of a better way to do this--the interactions with TangibleNet aligned with the visual effects. 
The magnets are physical, and the visualization is digital, but it felt seamless, unlike using a mouse or keyboard for presentations.''} Moreover, P5 observed that \textit{``Changing the node size by rotating the magnet is just like adjusting the volume on my stereo.''} 
Similarly, P4 commented, \textit{``Rotating the magnet to reveal the child network feels like zooming in with a DSLR camera lens to see more detail. I enjoy the sensation of diving into the child network.''} This suggests that TangibleNet effectively leverages the affordances of magnets and users' familiarity with physical objects. 
\ac{However, P11 remarked, \textit{``It feels inconsistent that physical objects represent nodes while the links remain non-physical.''} This inconsistency in aligning physical and digital elements affects the perceived naturalness of the interaction for some users.}

Some participants highlighted the drawbacks of body-based interactions. \ac{P1 and P11} mentioned experiencing arm fatigue during extended use and concerns about blocking the projected image, potentially affecting the audience's experience. Additionally, P1, P11, and P12 preferred directly manipulating projected visuals by hand for certain interaction commands. As P12 remarked, \textit{``When I want to group three nodes at once, I'd rather circle them with my finger on a touch display.''} 
Similarly, \rc{P1 stated, \textit{``It's easier to sketch a link directly on the whiteboard.''}}
\ac{P11 further added, \textit{``Not being able to interact with the visuals by touch makes some actions feel disconnected and less straightforward.''}}
These observations suggest that integrating TangibleNet with touch displays could enhance both the intuitiveness and flexibility of its interaction design.

\noindent \textbf{Mixed Reactions in Aligning Deictic Gestures and Interaction Commands:}
The design of aligning interaction commands with deictic gestures received mixed feedback. 
\rc{P9 appreciated this approach, stating, \textit{``Using gestures to display nodes and links helps to highlight points of interest.''}
They also remarked, \textit{``Compared to using a clicker, where I have to manage both the clicker and pointing at the same time, this system makes it easier since I can do both in one motion.''}
In contrast, P3 found it frustrating, explaining \textit{``I was pointing at a node to direct the audience's attention, and suddenly the annotation popped up. I didn't want that to happen.''} Reflecting on the experience, they added, \textit{``It's confusing when the same gesture is used for guiding the audience and controlling the visuals. It makes the software harder to use.''}}
This comment aligns with discussions on the conflict between operational and affective gestures, as noted by Hall\etal~\cite{Chironomia}. 
The issue is further exacerbated by the limitations of the current CV-based system through webcam, which lacks depth detection. 
\rc{This can result in pointing gestures being misinterpreted as interactions with the magnets, leading to unintended actions.}

\noindent \textbf{Benefits of Physical Interaction: }
\ac{Most} participants highlighted that using physical objects made interactions smoother and easier for presenters. 
\rc{P12 remarked, \textit{``This tool lets me multitask more efficiently than a touchscreen because I can use multiple magnets at the same time.''}
P2 added, \textit{``The magnets are thick enough to grab easily, even from the sides of the whiteboard. It's easier to handle physical objects than digital interfaces.''}}
\ac{These comments highlight how tangibility can make interactions more easy to use and efficient in data storytelling~\cite{TangibleBit, TUItextbook}.}
In addition, several participants appreciated the magnetic force.
\rc{P2 reflected, \textit{``The magnets stayed in place during rotations and slides, which made everything feel smooth and satisfying.''}
Others mentioned that the magnetic force added a tactile element that made the interactions more enjoyable. 
The familiarity of magnets further contributed to their ease of use.
P3 shared a nostalgic observation: \textit{``Using magnets in the presentation reminded me of arranging character magnets on a wall as a kid to represent relationships.''} These comments underline the value of physicality and its ability to make interactions easy to use and engaging.}

\noindent \textbf{Enhancing the Presenter's Sense of Autonomy: }
Participants frequently highlighted that TangibleNet \ac{provided a sense of autonomy during presentations, making the experience engaging.}
\ac{This was reflected in the responses, with 12 out of 14 participants describing their experience with TangibleNet as engaging (\cref{fig:experiment}-c).} 
\rc{P8 explained, \textit{``\ac{(In contrast to slideshow)} I liked being able to control when nodes and links appeared and deciding which elements to show. It gave me a sense of control over the presentation.'' } 
P2 also commented, \textit{``The magnets' weight and the way they stick to the board made it feel very hands-on. It reminded me of driving a manual car. I feel more connected to what I'm doing.''}
These remarks highlight how TangibleNet's physical interactions allowed presenters to actively construct network visualizations rather than simply advancing slides, contributing to a sense of autonomy and engagement.}

\noindent \textbf{Comparison with Existing Workflows: }
\rc{The instructors asked participants about the tools they typically use for data storytelling and how TangibleNet compares in terms of capabilities. Participants mainly use PowerPoint or Keynote slide decks, printed diagrams, and sketches for their presentations. They identified two key advantages of TangibleNet: flexibility and dynamic layout.
Many participants appreciated how TangibleNet enables more improvisational presentations. }
P12 noted, \textit{``Unlike slideshows with fixed sequences, this system lets me freely interact with the visualization and adapt the narrative during the presentation.''} This flexibility was particularly beneficial in dynamic environments, such as business settings where priorities and time constraints can shift rapidly. Additionally, the dynamic layout afforded by TangibleNet allowed participants to easily reposition nodes, optimizing space usage and bringing key components into focus as the story evolved. P12 further highlighted, \textit{``I don't have to pre-plan how the network visualization will change in the story.''} Observations confirmed that participants frequently adjusted the layout to reduce visual clutter as the network expanded.

However, participants noted some limitations. Several preferred alternative implementations of certain commands and requested advanced features to better support their storytelling needs. 
\rc{P5 mentioned, \textit{``Flipping magnets one by one to change node states doesn't work for scenarios like soccer, where players often have multiple roles.''}}
The current implementation, which only allows sequential state changes, felt restrictive in such contexts. 
\rc{P7 commented, \textit{``I wish I could change the state of all nodes in a category at the same time. It'd be helpful for things like showing character changes in a TV drama map after a major event.''}}
\rc{This feature was seen as particularly useful in scenarios where multiple node states shift simultaneously.}

\noindent \textbf{Feedback on Setup: }
Many participants appreciated TangibleNet's accessible setup, which uses standard equipment, making it easy to install in homes or offices. However, they noted challenges with the CV-based approach. P7, P8, and P10 struggled with interactions a few times when their bodies unintentionally blocked the markers from the webcam. Participants were sometimes unaware of the camera's role in tracking. 
\rc{P7 remarked, \textit{``I didn't realize my body was blocking the markers from the webcam until the interaction didn't work.''}
Similarly, P8 noted, \textit{``When I focused on using the magnets and telling the story, I forgot the camera was even there.''}
These comments reflect a lack of awareness about the camera's role in tracking interactions, particularly among participants like P7, P8, and P10, who had no prior experience with AR. }
Notably, P7, P8, and P10 had no experience with AR. 
\rc{These challenges suggest that users unfamiliar with AR and CV may require additional guidance to fully utilize TangibleNet.}
\section{Discussion}
We summarize key lessons and design implications for developing presentation tools with physical objects for synchronous data storytelling. We also discuss potential improvements.
\subsection{Design Implications}
\noindent \textbf{Expressiveness of User Actions:}
To ensure a smooth interaction experience, user actions should offer sufficient parameters for intended visualization commands \cite{PaperVis}. For example, repositioning one node offers two-dimensional continuous inputs, while flipping one node involves binary input. 
The dimension of object multiplexity in our design space allows the same user actions to generate different visualization commands based on timing and order, adding an extra layer of expressiveness. 
By incorporating physical objects, our prototype enhances the expressiveness of body language and gestures.
Currently, our prototype utilizes double-sided magnets as the primary interactive tool for network visualizations. 
However, other physical objects designed particularly for storytelling (e.g., Hans Rosling's meter-long teaching stick~\cite{rosling2015ignorant}) could progress stories in creative ways.

\noindent \textbf{Usability from the Presenter's Perspective:}
Usability in synchronous storytelling relies on how \rc{easy} the interactions are for the presenter, allowing them to concentrate on delivering the narrative smoothly. 
\rc{This is particularly important} compared to pre-recorded presentations, as synchronous storytelling requires quick thinking and improvisation~\cite{FromJamtoRecital}.
A key advantage of our prototype is its use of physical objects for interaction, leveraging their inherent \rc{familiarity} and simplicity. 
Most participants quickly learned how to interact with these objects by drawing on their real-world experiences and found the mapping between actions and interaction commands \rc{straightforward}.
\rc{Furthermore, interacting with physical objects enhances user autonomy and makes the storytelling experience more engaging. This increased engagement through physical interaction aligns with the existing research findings~\cite{tangibelGoodEngagement, Coda}. Additionally, allowing presenters to incrementally build visuals based on their judgment during the presentation further reinforces their sense of autonomy.}
\ac{We recommend that future designs incorporate physical interactions to simplify user interactions and enhance both presenter autonomy and engagement.}

\noindent \textbf{Engagement from the Audience's Perspective:}
While some interactions feel intuitive to the presenter, they might appear awkward or confusing to the audience. 
Neglecting this balance can divert attention from content and reduce engagement, a key concern frequently emphasized by news anchors trained in effective communication. 
\ac{Considering the audience's perspective on interactions, rather than focusing solely on the presenter, is especially critical in synchronous data storytelling. This requirement is distinct from interaction design for data analysis, where the observer's viewpoint typically receives minimal attention. As we learned from insights provided by news anchors, designing interaction commands for visualizations that leverage principles of engaging gestures in public communication is a necessary direction for future research, aiming to create interactions that are both functional for presenters and captivating for audiences.}


\noindent \textbf{Balancing Expressiveness, Usability, and Audience Engagement:}
Balancing expressiveness, usability, and audience engagement was challenging during prototyping. For example, as discussed in \cref{codesign}, the toggle action for hiding \rc{links} by bringing magnets closer together is \rc{straightforward} for the presenter but appears \rc{counterintuitive} to the audience because the \rc{links} disappear as the nodes move closer. This approach only meets the criteria for expressiveness and usability. Similarly, while the flip action for changing node states is easy to perform and does not seem \rc{counterintuitive}, it lacks the expressiveness needed to handle multiple states, addressing only usability and audience engagement. Future designs should consider integrating physical object-based interactions with multimodal inputs, such as speech or gaze, and a broader range of gestures to address all three criteria better.

\noindent \textbf{Automation vs. Manual Control: }
We recommend differentiating between elements the presenter controls manually and automated ones, depending on their significance in the narrative. 
\rc{Our findings suggest that manual control enhances the presenter's sense of autonomy and engagement, bringing flexibility to the storytelling.}
However, relying solely on manual control can be time-consuming and burdensome and lead to fatigue, particularly in lengthy or complex presentations. To address this, we implemented rules where specific nodes and \rc{links} appear or disappear based on changes in other components. The balance between manual and automatic control should be guided by the importance of the various story elements.
Additionally, incorporating live speech recognition for controlling visuals during the presenter's narration could help reduce their workload and facilitate a smoother storytelling experience, \ac{as demonstrated in prior synchronous storytelling research~\cite{RealityTalk}.}
\subsection{Limitations and Future Work}
\noindent\textbf{Expanding Network Storytelling}: 
Our interaction commands do not fully encompass those required to address a broader range of narrative patterns~\cite{SegelNarrative} and visual representations~\cite{bashDynamicGraphComic, kimDataToon}. For example, as P7 noted, our prototype lacks a command for filtering nodes by attributes.
To enhance TangibleNet's versatility, 
future work can explore multimodal inputs (e.g., speech, eye gaze), touch-enabled displays like smartboards, and various physical objects. Several participants expressed interest in these technologies to expand interaction capabilities.

\noindent\textbf{Lack of Authoring Support}:
This study does not address authoring support. While TangibleNet reduces the effort required for storytelling by enabling dynamic story construction through interaction, \rc{the initial network data setup still requires manual coding.}
Future work will develop a web-based interface to streamline this process, supporting authoring tasks
such as defining node states, adding annotations, previewing narratives, and configuring visual attributes like node appearance and animations. 
Its design will be guided by existing research on authoring systems for node-link diagrams for communications~\cite{ExpressiveAuthoringNodeLink, SpritzerStaticNet}.

\noindent\textbf{\ac{TangibleNet for Remote Audiences:}}
While TangibleNet was primarily designed for co-located storytelling with a whiteboard and projector, its core functionality---interacting with network data via physical objects---is also valuable for remote use.
\fc{TangibleNet can be adapted in two ways: (1) by streaming a recorded presentation, or (2) by integrating presenter tracking, similar to features in videotelephony software like Zoom and Google Meet~\cite{zoom,googlemeet}. This approach captures the presenter's body movements and physical interactions, embedding them into the video feed sent to remote users (e.g., \cite{Chironomia, RealityTalk, Elastica}). While this method can enhance visual clarity by eliminating the need to record a projected image, it may introduce potential errors or delays due to body detection and visual integration.}

\noindent\textbf{Exploring Other Physical Objects}: We chose double-sided magnets for their accessibility, familiarity, and ease of use. 
As their shape and properties resemble other objects, our design space may extend to similar physical items. However, other objects may offer unique affordances that enhance network visualization interactions, such as physical representations of links. Additionally, the semantic role of physical objects in storytelling is worth exploring. In theatre, props help convey narratives, suggesting that integrating such objects with visualizations could open new interaction possibilities for future research.

\noindent\textbf{Exploring Diverse Storytelling Practices}: 
Our formative study included five news anchors, experts in data communication--often-overlooked in visualization research. 
\fc{However, our evaluation did not involve news anchors or test the system in diverse settings like TV programs or classrooms. These contexts present unique challenges, such as varying audience sizes, audience interactions, and integration with other media (e.g., videos, music). Future research should evaluate the system in real-world environments with communication experts to address these challenges.}
\section{Conclusion}
This study explored interaction design with physical objects for network visualization in synchronous data storytelling. We interviewed five news anchors to identify key communication factors and the role of physical objects in presentations. We then conducted workshops with 14 VIS/HCI researchers to examine how physical objects can interact with network visualizations. These insights informed a three-dimensional design space: 1) interaction commands, 2) primary modality, and 3) multiplexity of physical objects. We developed TangibleNet, a projector-based AR prototype that allows presenters to interact with node-link diagrams using double-sided magnets. Our evaluation with 12 participants showed that TangibleNet supports \rc{interactions that are easy to learn}, enhances presenter autonomy and effectively supports synchronous data storytelling. We hope this work inspires future research on physical objects in data-driven storytelling.

\begin{acks}
We would like to thank the reviewers. This work is partially supported by the HK RGC GRF grant 16214623 and by the Knut and Alice Wallenberg Foundation through Grant KAW 2019.0024.
\end{acks}

\bibliographystyle{src/ACM-Reference-Format}
\bibliography{reference}

\end{document}